\documentclass[]{pasj01}
\Received{$\langle$reception date$\rangle$}
\Accepted{$\langle$acception date$\rangle$}
\usepackage{subfigure}
\usepackage{lineno}
\usepackage{ulem}

\begin{document}
%%%Title%%%%%%%%%%%%%%%%%%%%%%%%%%%%%%%%%%
\title{Testing the non-circularity of the spacetime around Sagittarius A* with orbiting pulsars}
%%%Authors%%%%
\author{
Yohsuke \textsc{Takamori}, \altaffilmark{1}$^{*}$
Atsushi \textsc{Naruko},\altaffilmark{2}
Yusuke \textsc{Sakurai},\altaffilmark{3}
Keitaro \textsc{Takahashi},\altaffilmark{4,5,6}
Daisuke \textsc{Yamauchi},\altaffilmark{7} and
Chul-Moon \textsc{Yoo}\altaffilmark{3}}
\altaffiltext{1}{National Institute of Technology (KOSEN), Wakayama College, Gobo, Wakayama 
644-0023, Japan}
\altaffiltext{2}{Center for Gravitational Physics,Yukawa Institute for Theoretical Physics,
Kyoto University, Kyoto 606-8502, Japan}
\altaffiltext{3}{Division of Particle and Astrophysical Science, Graduate School of Science,
Nagoya University, Nagoya 464-8602, Japan}
\altaffiltext{4}{Kumamoto University, Graduate School of Science
and Technology, Kumamoto, 860-8555, Japan}
\altaffiltext{5}{International Research Organization for Advanced Science and Technology,
Kumamoto University, Kumamoto 860-8555, Japan}
\altaffiltext{6}{National Astronomical Observatory of Japan,
2-21-1 Osawa, Mitaka, Tokyo 181-8588, Japan}
\altaffiltext{7}{Faculty of Engineering, Kanagawa University,
Kanagawa-ku, Yokohama-shi, Kanagawa, 221-8686, Japan}

\email{takamori@wakayama-nct.ac.jp}
%%%%%%%%%%%%%%%%%%%%%%%%%%%%%%%%%%%%%%%
\KeyWords{Galaxy: center --- pulsars: individual (orbiting pulsars) --- black hole physics}

\maketitle

%;Abstract%%
\begin{abstract}
A disformal Kerr black hole solution is a rotating black hole solution in a modified 
gravity theory which breaks the circular condition of spacetime differently from the case of the Kerr spacetime. 
In this paper, assuming that Sagittarius A* (Sgr A*) is a disformal Kerr black hole, 
we examine the potential to test the spacetime geometry with a hypothetical pulsar whose 
orbital elements are similar to those of the S2/S0-2 star. 
By numerically solving the equations of motion for the pulsar and 
photons emitted from it, 
we calculate the apparent position of the pulsar and the time 
of arrival (TOA) of the emitted pulse signals. 
Our analysis shows that the magnitude of the difference in the TOAs reaches 
the order of $10\>{\rm ms}$ if the deviation from the Kerr spacetime is significant. 
The time difference is mainly caused by the non-circularity of the spacetime at the $1.5$ 
post-Newtonian order. The accuracy of the TOA measurement by 
a future radio telescope named the Square Kilometer Array (SKA)
is between about $0.1\>{\rm ms}$ and $10\>{\rm ms}$ for a normal pulsar.
Thus, we expect that the SKA can distinguish the disformal Kerr black hole  
from the Kerr black hole through the non-circularity of the spacetime around Sgr A*.
\end{abstract}
%\linenumbers

\section{Introduction}
The existence of black holes in our universe is a general prediction of general relativity, 
and the black holes are crucial in many aspects of cosmology and astrophysics.
Astrophysical black holes can be classified by their mass, 
which is distributed in the wide range from about $1M_{\odot}$ to $10^9M_{\odot}$.
In general relativity, assuming vacuum environment, we can represent 
every black hole as the Kerr black hole thanks to the uniqueness theorem \citep{1975PhRvL..34..905R}.
In the 2010s, a number of exciting observational discoveries about the black holes 
have been made.
In 2015, the first detection of gravitational waves from a binary black hole has
been achieved by the advanced Laser Interferometer Gravitational-Wave Observatory (aLIGO) 
and the Virgo \citep{2016PhRvL.116f1102A}.
Then, in 2018, the general relativistic effects in the dynamics of 
the star orbiting Sagittarius A* (Sgr A*) have been detected 
by the Very Large Telescope (VLT), the Keck telescope, 
and the Subaru telescope \citep{2018AA...615L..15G, 2019Sci...365..664D}. 
Moreover, in 2019, the first image of the supermassive black hole in M87  
has been successfully obtained by the Event Horizon Telescope (EHT) \citep{2019ApJ...875L...1E}.
Those observational results are consistent with the Kerr black hole within their observational 
uncertainties.

One obviously attractive subject in observational research of black hole physics 
is the detection of the correction in gravitational theory beyond general relativity.
Once obtaining a black hole solution in a modified gravity theory, it can be used 
as an alternative to the Kerr black hole.
Deviation from the Kerr black hole implies modifications of the gravitational theory,
and therefore we can test the validity of general relativity through the test of the Kerr black hole.
Recently, a deformed Kerr black hole solution has been constructed 
in the context of the Degenerate Higher Order Scalar-Tensor (DHOST) theory 
[see the recent reviews of the DHOST theory:
 \citet{2019RPPh...82h6901K,2019IJMPD..2842006L}], 
which is called the disformal Kerr black hole \citep{2020JCAP...11..001B,2021JHEP...01..018A}.
The disformal Kerr black hole breaks the circular condition of spacetime, and
the non-circularity of spacetime can be regarded as a deviation 
from general relativity \citep{2020PhRvD.102h4021N,2021PhRvL.126x1104X}.
For the disformal Kerr spacetime, the non-circularity is characterized 
by a constant parameter $D$ called the disformal parameter.
When $D=0$, the disformal Kerr metric reduces to the Kerr metric, 
and therefore the non-vanishing value of the disformal parameter $D$ 
can be treated as a possible signal of the correction in the gravitational theory 
beyond general relativity.
Chen, Wang, and Jing (\yearcite{2021arXiv210311788C}) have applied 
the disformal Kerr black hole to GRO J1655-40 and analyzed 
its observational data. As a result, they have obtained the disformal parameter  
as $D=-0.010^{+0.011}_{-0.012}$ for GRO J1655-40. 
Moreover, the post-Newtonian (PN) motion of stars orbiting the disformal Kerr black hole 
has been analyzed using
the osculating orbit method \citep{2021PhRvD.103l4035A},
and it has been found that the secular evolution appears at $2$PN order in generic $D$.
When $D\sim -1$, the secular shift due to the disformal parameter becomes comparable
to the pericenter shift due to the Schwarzschild potential.
In 2018, S2/S0-2 (hereafter S2) that is an orbiting star around Sgr A*
passed through its pericenter and 
the pericenter shift is consistent with general relativity \citep{2020AA...636L...5G}.
This observation of S2 would not allow the disformal Kerr black hole with $D\sim -1$ 
to represent Sgr A*. 

This paper analyzes the dynamics of hypothetical pulsars orbiting Sgr A* assuming that 
Sgr A* can be well described by the disformal Kerr black hole with $D>-1$.
Orbiting pulsars are suitable probes to study the spacetime around 
compact objects thanks to their accurate pulse period
\citep{1999ApJ...514..388W, 2006Sci...314...97K, 2012ApJ...747....1L,2016ApJ...818..121P}.
In the Galactic Center region, orbiting stars within 
$0.02\>{\rm pc}$ have been discovered [see the review \citet{2010RvMP...82.3121G}].
Some of them are young and massive, and they can leave a neutron star at the end of life. 
Some of the neutron stars would be observed as a normal pulsar
and not only main sequence stars but also pulsars could be orbiting Sgr A*
\citep{2004ApJ...615..253P,2012ApJ...753..108W}.
Although no normal pulsars have been detected within the inner parsec yet 
\citep{2009ApJ...702L.177D,2010ApJ...715..939M,2011MNRAS.411.1575B,
2021ApJ...914...30L,2021AA...650A..95T},
a magnetar whose distance from Sgr A* is $\sim 0.1\>{\rm pc}$ 
was discovered \citep{2013Natur.501..391E, 2013ApJ...770L..24K, 2013ApJ...770L..23M}.
In 2021, construction of a new facility named the Square Kilometer Array (SKA) has begun, 
which will be the largest radio telescope in the world.
The SKA would detect about 14,000 normal pulsars \citep{2009AA...493.1161S}, 
and some of them would be orbiting Sgr A*.
The dynamics of a hypothetical pulsar on the S2-like orbit has been 
investigated with Sgr A* being the Kerr black hole \citep{2017ApJ...849...33Z}.
The spin of the central black hole 
is detectable by continuous monitoring of the hypothetical pulsar with the SKA.
In this paper, we conduct a similar analysis by using the disformal 
Kerr black hole solution.
Investigating a hypothetical orbiting pulsar with the orbital 
elements similar to S2, we find that
the deviation from the Kerr in the arrival time of the pulse 
mainly comes from the non-circularity of the disformal Kerr at the $1.5$PN order.
Moreover, we discuss the detectability of the disformal parameter $D$
by observations of the hypothetical pulsar with the SKA.

This paper is organized as follows. 
In Sec.~\ref{sec:dKerr}, we introduce the disformal Kerr solution.
In Sec.~\ref{sec:orbitingPulsar}, we show our method to test the non-circularity of 
the spacetime around Sgr A* with hypothetical orbiting pulsars.
The observables are the pulsar's position on the sky and the arrival time of the pulse.
Those are determined by the motion of the pulsar and emitted photons, 
and the motion contains the effects of the non-circularity of spacetime.
We numerically solve the equations of motion with the Hamiltonian formalism 
and demonstrate the effects of the disformal parameter.
In Sec.~\ref{sec:results}, we show results obtained by our simulation and
discuss the detectability of $D$ with the SKA.
Finally, we summarize our work in the last section.
Throughout this paper, we set $c=G=1$ where $c$ and $G$ are the speed of light and 
the gravitational constant, respectively.

\section{Disformal Kerr solution}
\label{sec:dKerr}
We summarize a derivation of the disformal Kerr solution 
\citep{2020JCAP...11..001B, 2021JHEP...01..018A} and its properties.
Let us consider the action of a scalar field with its higher derivatives:
\begin{eqnarray}
  &&S=\int d^4x\sqrt{-g}
  \Biggl{[}P(X, \phi)+Q(X, \phi)\Box\phi+F(X, \phi)R \nonumber \\
  &&\hspace{5cm}+\sum_{i=1}^5A_{i}(X, \phi)L_{i} \Biggr{]},
\end{eqnarray}
where $P$, $Q$, $F$, and $A_{i}$ are functions depending on the scalar field $\phi$ and 
its kinetic term $X=\nabla_{\mu}\phi\nabla^{\mu}\phi$. 
$\nabla_{\mu}$ is the covariant derivative associated with the metric $g_{\mu\nu}$, 
and $R$ is the four-dimensional Ricci scalar. $L_i$ are given by
\begin{eqnarray} &&L_1=\phi_{\mu\nu}\phi^{\mu\nu},~~L_2=(\Box\phi)^2,~~L_3=\phi^\mu\phi_{\mu\nu}\Box\phi, 
  \nonumber \\
 &&L_4=\phi^{\mu}\phi_{\mu\nu}\phi^{\nu\rho}\phi_{\rho},
    ~~L_5=\left(\phi^\mu\phi_{\mu\nu}\phi^{\nu}\right)^2,
\end{eqnarray}
where $\phi_{\mu}=\nabla_{\mu}\phi$ and $\phi_{\mu\nu}=\nabla_{\nu}\nabla_{\mu}\phi$.
In the DHOST theory, $F$ and $A_i$ are not free functions, 
and satisfy the specific conditions which guarantee 
the absence of ghost instabilities associated with the higher derivatives 
\citep{2016PhRvD..93l4005B, 2016JCAP...02..034L}. 
In this paper, we focus on a subclass of DHOST theories, called the type-I DHOST theory. 
It is known that a DHOST theory with a set of free functions can be mapped to a different DHOST theory with another set of functions through a conformal-disformal transformation.  
The transformed metric $\tilde g_{\mu\nu}$ is
expressed in terms of the original metric $g_{\mu\nu}$
and the scalar field $\phi$ as
\begin{equation}
  \tilde{g}_{\mu\nu}={\cal A}(X, \phi)g_{\mu\nu}+{\cal B}(X,\phi)\partial_{\mu}\phi\partial_{\nu}\phi\,.
\end{equation}
This fact enables us to construct a new solution in a DHOST theory from an existing solution in 
another specific DHOST theory by performing a metric transformation.

Recently, a rotating black hole solution in a (subclass of type-I) DHOST theory
has been found \citep{2019PhRvD.100h4020C}.
Notably, its geometry is described by the Kerr solution 
while the scalar field has a nontrivial profile, and therefore 
the solution is called the stealth Kerr solution.
The metric of the Kerr spacetime with a mass $M$ and a spin $a$ is written in the Boyer-Lindquist coordinates, $(t, \varphi, r, \theta)$, as
\begin{eqnarray}
  g_{\mu\nu}^{\rm Kerr}dx^{\mu}dx^{\nu} &=&-\left(1-\frac{2Mr}{\rho^2}\right)dt^2
  -\frac{4Mar\sin^2\theta}{\rho^2}dtd\varphi \nonumber \\
  &&\hspace{1.2cm}+\frac{A\sin^2\theta}{\rho^2}d\varphi^2 
  +\frac{\rho^2}{\Delta}dr^2+\rho^2 d\theta^2,
\end{eqnarray} 
where 
\begin{eqnarray}
 \Delta &=& r^2-2Mr+a^2, \\
 \rho^2 &=& r^2+a^2\cos^2\theta,\\
 A &=& (r^2+a^2)^2-a^2\Delta\sin^2\theta\,.
\end{eqnarray}
The spin parameter is restricted in the range $0\leq |a/M| \leq 1$ for a black hole spacetime.
The scalar field solution with a nontrivial configuration is given by
\begin{equation}
\phi=q\left[t+\int\frac{\sqrt{2Mr(a^2+r^2)}}{r^2-2Mr+a^2}dr\right],
\end{equation}
where $q$ is a positive constant. 

Applying a disformal transformation to the stealth Kerr solution, a new solution 
can be generated as
\begin{equation}
  \tilde{g}_{\mu\nu}^{\rm Kerr}
  =g_{\mu\nu}^{\rm Kerr}-\frac{D}{q^2}\partial_\mu\phi\partial_\nu\phi,
  \label{eq:dtrans}
\end{equation}
where $D$ is a constant called the disformal parameter. 
The set of $(\tilde{g}_{\mu\nu}^{\rm Kerr}, \phi)$ is also an exact solution 
in another subclass of DHOST theories which includes $D$ 
as a parameter in the gravity theory.
From equation (\ref{eq:dtrans}), we see that 
$\tilde{g}^{\rm Kerr}_{tt}\sim -(1+D)$ in the asymptotic region $r\gg M$. 
Thus, the disformal parameter $D$ should 
satisfy the inequality $D>-1$ to keep the metric signature $(-,\,+,\,+,\,+)$ 
for $\tilde{g}^{\rm Kerr}_{\mu\nu}$.
Introducing the new time coordinate by $t \rightarrow t/\sqrt{1+D}$ 
in equation (\ref{eq:dtrans}), namely 
$\tilde{g}_{\mu\nu}^{\rm Kerr} [t/\sqrt{1+D}, \varphi, r, \theta]=
g_{\mu\nu}^{\rm dKerr} [t, \varphi, r, \theta]$, 
the metric of the disformal Kerr solution reads
\begin{eqnarray}
  &&g_{\mu\nu}^{\rm dKerr}dx^{\mu}dx^{\nu} \nonumber \\
  &&=-\left(1-\frac{2\tilde{M}r}{\rho^2}\right)dt^2
-\frac{4\sqrt{1+D}\tilde{M}ar\sin^2\theta}{\rho^2}dtd\varphi \nonumber \\
&&~~~~+\frac{A\sin^2\theta}{\rho^2}d\varphi^2
+\frac{\rho^2\Delta-2\tilde{M}(1+D)Dr(a^2+r^2)}{\Delta^2}dr^2  \nonumber \\
&&\hspace{2.7cm}-2D\frac{\sqrt{2\tilde{M}r(a^2+r^2)}}{\Delta}dtdr
+\rho^2d\theta^2\,.
   \label{metric:dKerr}
\end{eqnarray}
Here we have introduced the rescaled mass $\tilde{M}=M/(1+D)$.
From equation (\ref{metric:dKerr}), we see that the mass and spin 
measured by a distant observer are given by
$\tilde{M}$ and $\tilde{a}=\sqrt{1+D}a$ not by $M$ and $a$. 

The most significant difference between the Kerr and the disformal Kerr is 
whether the cross term $g_{tr}$ exists.
In general relativity, the Kerr spacetime satisfies the circular condition
which roughly implies the absence of cross terms between $(t, \varphi)$ and $(r, \theta)$.
For $D=0$, the cross term $g_{tr}$ becomes zero, 
and then equation (\ref{metric:dKerr}) gives the Kerr metric
in the Boyer-Lindquist coordinates.
In the static limit that $a=0$, non-zero $g_{tr}$ remains, 
but it can be removed by the following coordinate transformation \citep{2021JHEP...01..018A}:
\begin{equation}
  dt=dT-\frac{D\sqrt{2\tilde{M}r^3}}{\Delta\left(1-\frac{2\tilde{M}}{r}\right)}dr\,.
\end{equation}
Then, the metric is written by
\begin{equation}  
  ds^2
=-\left(1-\frac{2\tilde{M}}{r}\right)dT^{2}+\frac{dr^2}{1-\frac{2\tilde{M}}{r}} +r^2d\theta^2
+r^2\sin^2\theta d\varphi^2.
\end{equation}
This metric represents the Schwarzschild spacetime with the rescaled mass $\tilde{M}$.
Hence, the non-circularity of the disformal Kerr spacetime appears with the spin parameter.

The condition that the disformal Kerr spacetime has the event horizon 
constrains the parameters $a$ and $D$ \citep{2021JHEP...01..018A}.
In the case $D\geq 0$, the critical value $a_{\rm c}$ is obtained by
\begin{equation}
  a_{\rm c}=\frac{\tilde{M}}{\sqrt{1+4D}}\,. 
\label{critical_a}
\end{equation}
The spin parameter $a$ should be in the range $|a|\leq a_{\rm c}$ for the disformal Kerr 
spacetime to represent a black hole spacetime.
In the case $-1<D<0$, one can obtain a polynomial equation for $a_{\rm c}$ 
[see Appendix B in \citet{2021JHEP...01..018A}].
The polynomial equation gives a finite value of $a_{\rm c}$ in the range $-1<D<0$. 
Taking the limit $D\rightarrow-1$, 
we obtain the critical value as $a_{\rm c}\sim0.516\tilde{M}$.
Recall that the spin of the disformal Kerr is not $a$ but $\tilde{a}=\sqrt{1+D}a$. 
In terms of $\tilde{a}$, the critical value for $D\geq0$ is expressed as 
\begin{equation}
  \tilde{a}_{\rm c}=\tilde{M}\sqrt{\frac{1+D}{1+4D}}.
\label{critical_tilde_a}
\end{equation}
For $D=0$, which is the Kerr case, we have $\tilde{a}_{\rm c}=\tilde{M}$.
Taking the limit $D\rightarrow\infty$, the maximum spin becomes $\tilde{M}/2$.
In the range $-1\leq D<0$,  
$\tilde{a}_{\rm c}$ approaches zero in the limit $D\rightarrow-1$ because 
$a_{\rm c}\sim 0.516\tilde{M}$.
This study focuses on the parameter range for which the event horizon exists.

\section{Method to test the non-circularity of the spacetime around 
Sgr A* with orbiting pulsars}
\label{sec:orbitingPulsar}
Assuming that Sgr A* is a disformal Kerr black hole, 
we examine the motion of hypothetical pulsars on the orbit like S2.
We call a pulsar whose orbital elements are similar to those of S2
a S2-like pulsar hereafter following Zhang and Saha (\yearcite{2017ApJ...849...33Z}).
We also assume that the standard matter fields including the pulsar and photons 
are minimally coupled with the disformal metric $g_{\mu \nu}^{\rm dKerr}$.
The ratio of the mass of the pulsar to that of Sgr A* is about $10^{-6}$.
Thus, the pulsar can be regarded as a test particle and follows a geodesic motion 
in the disformal Kerr spacetime.
In table \ref{tab:param}, we summarize the following parameters:
the mass of the black hole, the distance to the Galactic Center, 
and the typical orbital parameters of a S2-like pulsar.

\begin{table}[htbp]
  \caption{Summary of parameters.} \label{tab:param}
  \vspace{0.1cm}
  \begin{tabular}{ccc}\hline     
    \hspace{-0.2cm}Parameter & \hspace{-0.2cm}Description & \hspace{-0.2cm}Value \\
    \hline 
    \hspace{-0.2cm}$M_{\rm BH}\>(10^6M_{\odot})$     & \hspace{-0.2cm}{\footnotesize Black Hole Mass} & \hspace{-0.2cm}4.0\\
    \hspace{-0.2cm}$R_{0}\>({\rm kpc})$    & \hspace{-0.2cm}{\footnotesize Distance to the Galactic Center}& \hspace{-0.2cm}8.0\\
    \hline
    \hspace{-0.2cm}$e$ & \hspace{-0.2cm}{\footnotesize Eccentricity} &\hspace{-0.2cm}0.88\\
    \hspace{-0.2cm}$T\>({\rm yr})$&\hspace{-0.2cm}{\footnotesize Orbital Period} & \hspace{-0.2cm}16\\
    \hspace{-0.2cm}$I\>({\rm deg})$&\hspace{-0.2cm}{\footnotesize Inclination }&  \hspace{-0.2cm}135\\
    \hspace{-0.2cm}$\omega\>({\rm deg})$&\hspace{-0.2cm}{\footnotesize Argument of Periapsis}& \hspace{-0.2cm}65\\
    \hspace{-0.2cm} $\Omega\>({\rm deg})$&\hspace{-0.2cm}{\footnotesize Longitude of Ascending Node} & \hspace{-0.2cm}225\\ \hline
  \end{tabular}
\end{table}

\subsection{Coordinate transformation}
We examine the motion of the S2-like pulsar orbiting 
the disformal Kerr black hole for various $a$ and $D$.
Unfortunately, the original metric given in equation 
(\ref{metric:dKerr}) is unsuitable for our purpose.
Hence, we introduce a convenient coordinate system for our analysis below.

When $a=0$, equation (\ref{metric:dKerr}) fails to recover
the usual Schwarzschild metric due to the term $g_{tr}$.
Moreover, $g_{rr}$ in the metric obviously diverges when $\Delta=0$.
This singularity is the coordinate singularity of the metric. 
Solving the equation $\Delta=0$, we obtain the solutions for $r$
as a function of $\tilde{M}$ and $\tilde{a}$ as follows:
\begin{equation}
  \tilde{r}_{\rm d\pm}
  =(1+D)\tilde{M}\pm\sqrt{(1+D)^2\tilde{M}^2-\frac{\tilde{a}^2}{1+D}}\,. 
  \label{rd}
\end{equation}
The periapsis distance of the S2-like pulsar is about $3000M_{\rm BH}$.
When $D\sim1500$, the pulsar meets $\tilde{r}_{\rm d+}$ sometime.
Therefore, solving the equation of motion for the S2-like pulsar,
we face divergence when it passes through $\tilde{r}_{\rm d+}$.
New coordinates are given by the following coordinate 
transformation \citep{2021JHEP...01..018A}:
\begin{equation}
  dt
  =dT
  -D\frac{\sqrt{2\tilde{M}r(r^2+a^2)}}{\Delta\left(1-\frac{2\tilde{M}}{r}\right)}dr\,.
\end{equation}
With these new coordinates, the cross term $g_{Tr}$ is given by
\begin{equation}
  g_{Tr}
  =\frac{2D\tilde{M}a^2\cos^2\theta\sqrt{2\tilde{M}r(r^2+a^2)}}
  {(r-2\tilde{M})\rho^2\Delta}\,. \label{eq:gTr}
\end{equation}
The cross term $g_{Tr}$ vanishes when $a=0$, 
and the usual Schwarzschild metric comes back. 
However, the coordinate singularity where $\Delta=0$ still remains in equation (\ref{eq:gTr}). 
Therefore, that coordinate transformation is unsuitable for our purpose.

We shall consider a new coordinate system that reproduces
the usual Schwarzschild metric without the coordinate singularity at $\Delta=0$.
Let us perform the following coordinate transformations to
equation (\ref{metric:dKerr}):
\begin{eqnarray}
   dt&=&
         d\tilde{t}-\frac{D\sqrt{2\tilde{M}r(r^2+a^2)^3}}{\tilde{\Delta}\Delta}dr,\\
  d\varphi&=&
              d\tilde{\varphi}-
              \frac{D a \sqrt{2\tilde{M}r(r^2+a^2)}}{\sqrt{1+D}\tilde{\Delta}\Delta}dr,
\end{eqnarray}
where $\tilde{\Delta}=r^2-2\tilde{M}r+a^2$. 
Then, we have the metric with the new coordinates, $(\tilde{t},\tilde{\varphi},r,\theta)$, as 
\begin{eqnarray}
  &&\tilde{g}_{\mu\nu}^{\rm dKerr}dx^{\mu}dx^{\nu} \nonumber \\
 && =-\left(1-\frac{2\tilde{M}r}{\rho^2}\right)d\tilde{t}\,^2
     -\frac{4\sqrt{1+D}\tilde{M}ar\sin^2\theta}{\rho^2}d\tilde{t}d\tilde{\varphi} \nonumber \\
     &&~~~~+\frac{A\sin^2\theta}{\rho^2}d\tilde{\varphi}^2
  +\frac{\rho^2\tilde{\Delta}+D(r^2+a^2)(\tilde{\Delta}-a^2\sin^2\theta)}
   {(1+D)\tilde{\Delta}^2}dr^2 \nonumber \\
   &&\hspace{1.5cm}-\frac{2Da\sqrt{2\tilde{M}r(r^2+a^2)}\sin^2\theta}{\sqrt{1+D}\tilde{\Delta}}
   drd\tilde{\varphi}+\rho^2d\theta^2.
   \label{metric:dKerr2}
\end{eqnarray}
In this way, we can remove the coordinate singularity at $\Delta=0$. 
The cross term $g_{r\tilde{\varphi}}$ appears instead of $g_{tr}$ in the metric.
In the asymptotic region $r\gg\tilde{M}$,  equation (\ref{metric:dKerr2}) is approximated by
\begin{equation}
  \tilde{g}_{\mu\nu}^{\rm dKerr}dx^{\mu}dx^{\nu} 
  \sim -d\tilde{t}\,^2+r^2\sin^2\theta d\tilde{\varphi}^2+dr^2+r^2d\theta^2.
\end{equation}
Thus, the new coordinates represent the well-known polar coordinates 
in the Minkowski spacetime in the asymptotic region. 
Furthermore, equation (\ref{metric:dKerr2}) recovers 
the usual Schwarzschild metric when $a=0$, as expected.

We can also find that $g_{rr}$ and $g_{r\tilde{\varphi}}$ in equation (\ref{metric:dKerr2})
diverge when $\tilde{\Delta}=0$, and it is the coordinate singularity of the metric.
However, unlike the coordinate singularity at $\Delta=0$, 
as is shown below, the coordinate singularity at $\tilde{\Delta}=0$
is well inside the orbital radius of the S2-like pulsar, and we ignore
this singularity in this work. 
Solving the equation $\tilde{\Delta}=0$, we obtain the solutions as a function
of $\tilde{M}$ and $\tilde{a}$ as
\begin{equation}
  \tilde{r}_{\pm}=\tilde{M}\pm\sqrt{\tilde{M}^2-\frac{\tilde{a}^2}{1+D}}.
\end{equation}
In general, $\tilde{r}_+$ is grater than $\tilde{r}_{-}$ and outside the event horizon.
Moreover, we can see that $\tilde{r}_+\sim 2\tilde{M}$ for a generic $D$.
Because the orbital radius of the S2-like pulsar is always greater than $\tilde{r}_{+}$,
the pulsar will not meet the singularity at $r=\tilde{r}_{+}$.
Thus, the metric given in equation (\ref{metric:dKerr2}) is suitable for our purpose, and 
we solve the equations of motion for the S2-like pulsar and emitted photons with this metric.

\subsection{Numerical method}
This subsection shows our numerical method to solve the equations of motion 
for the S2-like pulsar and emitted photons.
It should be noted that we hereinafter suppress tildes in the metric 
as well as the coordinate for the sake of notational simplicity.

\subsubsection{Hamiltonian formalism}
This work uses the Hamiltonian formalism, which is 
convenient to deal with both the pulsar motion and a photon trajectory
 \citep{2010ApJ...711..157A}.
Letting $p_\mu$ denote the four-momentum of the pulsar or a photon, 
we have the Hamiltonian as follows:
\begin{equation}
  H=\frac{1}{2}g_{\rm dKerr}^{\mu\nu}p_{\mu}p_{\nu}
  =-\frac{\kappa}{2},
  \label{Hamiltonian}
\end{equation}
where $g_{\rm dKerr}^{\mu\nu}$ is the inverse of $g_{\mu\nu}^{\rm dKerr}$
and $\kappa$ is the constant given by $\kappa=1$ for the pulsar or $\kappa=0$ 
for a photon.
From equation (\ref{metric:dKerr2}), the components of $g^{\mu\nu}_{\rm dKerr}$
are given by
\begin{eqnarray}
  g^{tt}_{\rm dKerr}&=&
  -\frac{(r^2+a^2)^2-a^2\tilde{\Delta}\sin^2\theta}{\tilde{\Delta}\rho^2} \nonumber \\
 &&~~~~+\frac{4Da^2\tilde{M}^2r^2(r^2+a^2)\sin^2\theta}{\tilde{\Delta}^2\rho^4}, \\ 
  g^{tr}_{\rm dKerr} &=&
-\frac{2Da^2\sqrt{2\tilde{M}^3r^3(r^2+a^2)}\sin^2\theta}{\tilde{\Delta}\rho^4}, \\ 
  g^{t\varphi}_{\rm dKerr} &=&
  -\frac{2\sqrt{1+D}a\tilde{M}r}{\tilde{\Delta}\rho^2}
+\frac{4Da^3\tilde{M}^2r^2\sin^2\theta}{\sqrt{1+D}\tilde{\Delta}^2\rho^4}, \\ 
  g^{rr}_{\rm dKerr} &=&
\frac{\tilde{\Delta}}{\rho^2}+\frac{2Da^2\tilde{M}r\sin^2\theta}{\rho^4}, \\ 
  g^{r\varphi}_{\rm dKerr} &=&  
\frac{aD\sqrt{2\tilde{M}r(r^2+a^2)}(\tilde{\Delta}-a^2\sin^2\theta)}{\sqrt{1+D}\tilde{\Delta}\rho^4}, \\ 
  g^{\theta\theta}_{\rm dKerr} &=& \frac{1}{\rho^2}, \\ 
  g^{\varphi\varphi}_{\rm dKerr} &=&
\frac{\tilde{\Delta}-a^2\sin^2\theta}{\tilde{\Delta}\rho^2\sin^2\theta}
-\frac{2Da^2\tilde{M}r(\tilde{\Delta}-a^2\sin^2\theta)}{(1+D)\tilde{\Delta}^2\rho^4} \,.
\end{eqnarray}

When numerically solving the equation of motion for photons emitted from an orbiting star,
the pseudo-polar coordinates, $(\varphi, r, \theta)$,  
are unsuitable \citep{2010ApJ...711..157A}.
This is because the photon travels to a distant observer being 
$8\>{\rm kpc}\sim10^{10}M_{\rm BH}$ away from the black hole,
and the changes of the angular coordinates $(\varphi, \theta)$ 
are extremely small near the distant observer.
Therefore, we use Cartesian-like coordinates to 
solve Hamilton's equations for the photons.
This work simply uses the pseudo-Cartesian coordinates, $(x, y, z)$, defined by
\begin{equation}
  x = r\sin\theta\cos\varphi,~~y = r\sin\theta\sin\varphi,~~z = r\cos\theta\,.
\end{equation}
The pseudo-Cartesian coordinates represent the natural Cartesian 
coordinates in the asymptotic region $r\gg\tilde{M}$.
With the pseudo-Cartesian coordinates,
the spatial components of the four-momentum $p_{\mu}$ can be expressed 
as follows:
\begin{eqnarray}
  p_{r}&=&\frac{xp_x+yp_y+zp_z}{\sqrt{x^2+y^2+z^2}}, \\
  p_{\theta}&=&\frac{z(xp_x+yp_y)-(x^2+y^2)p_z}{\sqrt{x^2+y^2}}, \\
  p_{\varphi}&=& xp_y-yp_x.
\end{eqnarray}
When solving Hamilton's equations for the S2-like pulsar,
we also use the pseudo-Cartesian coordinates.
Finally, Hamilton's equations with the pseudo-Cartesian coordinates are written by
\begin{equation}
  \frac{dp_{\mu}}{d\lambda} = -\frac{\partial H}{\partial x^{\mu}},~~
  \frac{dx^{\mu}}{d\lambda} = \frac{\partial H}{\partial p_{\mu}}\,,
  \label{eq:HamiltonEquations}
\end{equation}
where $p_{\mu}=(p_{t}(\lambda), p_x(\lambda), p_y(\lambda), p_z(\lambda))$
and $x^{\mu}=(t(\lambda), x(\lambda), y(\lambda), z(\lambda))$
with $\lambda$ being an affine parameter.
To avoid confusion, we use $p_{\mu}$ for the pulsar and $k_{\mu}$ for 
an emitted photon hereafter.
For the pulsar, the affine parameter $\lambda$ can be regarded as 
its proper time, and we replace $\lambda$ with $\tau$.
To sum up, for the S2-like pulsar, we express Hamilton's equations as
\begin{equation}
  \frac{dp_{\mu}}{d\tau} = -\frac{\partial H}{\partial x^{\mu}},~~
  \frac{dx^{\mu}}{d\tau} = \frac{\partial H}{\partial p_{\mu}},
  \label{eq:HE_pul}
\end{equation}
with $H=-1/2$. For an emitted photon, we have
\begin{equation}
  \frac{dk_{\mu}}{d\lambda} = -\frac{\partial H}{\partial x^{\mu}},~~
  \frac{dx^{\mu}}{d\lambda} = \frac{\partial H}{\partial k_{\mu}},
  \label{eq:HE_ph}
\end{equation}
with $H=0$.

\subsubsection{Post-Newtonian and post-Minkowskian approximations}
The S2-like pulsar is orbiting well far from the central black hole.
Moreover, we do not consider photon trajectories passing through 
the vicinity of the central black hole. 
The post-Newtonian and post-Minkowskian approximations are
available in that case \citep{2010ApJ...711..157A}.
Although we can perform the same analysis with the exact expression of the Hamiltonian, 
the post-Newtonian and post-Minkowskian expansions are helpful to classify and understand 
the effects of the disformal parameter $D$.
We show the approximated Hamiltonian up to
$2$PN order because the effects up to $2$PN order, which include
the effects of the black hole spin and quadrupole moment, 
would be detectable in the SKA era.

For the S2-like pulsar, we have the following relation:
\begin{equation}
  v^2 \sim \frac{M_{\rm BH}}{r},
\end{equation}
where $v$ is the speed of the pulsar. Let $\epsilon$ be a small parameter satisfying 
$0<\epsilon\ll1$ and $v\sim \epsilon$.
Then, we can easily find that $r/M_{\rm BH} \sim 1/v^2\sim \epsilon^{-2}$.
The time-component of the four-momentum, $p_t$, is the order 
of unity because it gives the rest mass of the pulsar.
The $r$-component $p_r$ is the order of $\epsilon$ because $p_r\sim v$.
Moreover, $p_\theta/r \sim p_\varphi/r \sim \epsilon$ because $p_\theta/r \sim p_\varphi/r \sim v$.
Then, we can re-scale the four-momentum of the pulsar and $r$ by $\epsilon$ as follows:
\begin{equation}
  p_{t}\rightarrow p_{t},~p_{r}\rightarrow \epsilon p_{r},
  ~p_{\theta}\rightarrow \epsilon^{-1}p_{\theta},
  ~p_{\varphi}\rightarrow \epsilon^{-1}p_{\varphi},
  ~r \rightarrow \epsilon^{-2}r\,.
  \label{eq:rescale_pul}
\end{equation}
We substitute equation (\ref{eq:rescale_pul}) to equation (\ref{Hamiltonian})
and expand the Hamiltonian up to $2$PN order.
The approximated Hamiltonian for the pulsar is given by
\begin{eqnarray}  
 &&H \approx \frac{1}{2}\biggl{(} H^{\rm 0PN}_0+\epsilon^2H^{\rm 0PN}_2 
+\epsilon^4H^{\rm 1PN}_4 \nonumber \\
&&\hspace{4cm}+\epsilon^5H^{\rm 1.5PN}_5  +\epsilon^6H^{\rm 2PN}_6\biggr{)},
\end{eqnarray}
where
\begin{eqnarray}
  H^{\rm 0PN}_0&=&-p_{t}^2, \\ 
  H^{\rm 0PN}_2&=&
  -\frac{2\tilde{M}}{r}p_{t}^2
 +p_r^2+\frac{p_\theta^2}{r^2}+\frac{p_{\varphi}^2}{r^2\sin^2\theta}, \\ 
  H^{\rm 1PN}_4&=&-\frac{4\tilde{M}^2}{r^2}p_{t}^2-\frac{2\tilde{M}}{r}p_{r}^2, \\ 
  H^{\rm 1.5PN}_5&=&-\frac{4\tilde{a}\tilde{M}}{r^3}p_{t}p_{\varphi}
                  +\frac{2D\tilde{a}}{1+D}\sqrt{\frac{2\tilde{M}}{r^5}}p_{r}p_{\varphi}, \\ 
  H^{\rm 2PN}_6&=&
-\frac{8(1+D)\tilde{M}^3-2\tilde{a}^2\tilde{M}\cos^2\theta}{(1+D)r^3}p_{t}^2
+\frac{\tilde{a}^2\sin^2\theta}{(1+D)r^2}p_{r}^2 \nonumber \\
&&~~~~-\frac{\tilde{a}^2\cos^2\theta}{(1+D)r^4}p_{\theta}^2 -\frac{\tilde{a}^2}{(1+D)r^4\sin^2\theta}p_{\varphi}^2\,,
\end{eqnarray}
with $\tilde{a}=\sqrt{1+D}a$.
The post-Newtonian order was shown with superscripts.
$H^{\rm 0PN}_0$ represents the rest mass of the S2-like pulsar, and  
$H^{\rm 0PN}_2$ produces the Kepler motion.
$H^{\rm 1PN}_4$ is the weak-field Schwarzschild term that gives the pericenter shift.
$H^{\rm 1.5PN}_5$ gives the frame-dragging effect due to the spin of the black hole.
Then, the highest order term $H^{\rm 2PN}_6$ includes the quadrupole moment effects
of the black hole proportional to $\tilde{a}^2$.
We find that the leading correction due to the disformal parameter $D$ 
appears in $H^{\rm 1.5PN}_5$, and it gives the additional frame-dragging effect.

We can also expand the Hamiltonian for an emitted photon with a small 
parameter $\epsilon$, which is called the post-Minkowskian expansion.
We can assume that $r/M_{\rm BH}\sim \epsilon^{-2}$ as was 
in the case of the S2-like pulsar, 
but, for the four-momentum of the photon, the order is different from the pulsar. 
Because the photon travels with the speed of light, 
the four-momentum of the photon satisfies the relation
 $k_t\sim k_r \sim k_\theta/r\sim k_\varphi/r\sim 1$.
Thus, we re-scale $k_\mu$ and $r$ as follows:
\begin{equation}
  k_{t}\rightarrow k_{t},~k_{r}\rightarrow k_{r},
  ~k_{\theta}\rightarrow \epsilon^{-2}k_{\theta},
  ~k_{\varphi}\rightarrow \epsilon^{-2}k_{\varphi},
  ~r \rightarrow \epsilon^{-2}r.
  \label{eq:rescale_ph}
\end{equation}
By substituting equation (\ref{eq:rescale_ph}) to equation (\ref{Hamiltonian}), 
the Hamiltonian for the photon can be expanded up to the order of $\epsilon^6$ that
corresponds to $2$PN order:
\begin{eqnarray}  
H\approx \frac{1}{2}
\biggl{(}
&&H^{\rm 0PM}_0+\epsilon^2H^{\rm 1PM}_2\nonumber \\
&&\hspace{1.2cm}+\epsilon^3H^{\rm 1.5PM}_3+\epsilon^4H^{\rm 2PM}_4+\epsilon^6H^{\rm 3PM}_6\biggr{)},
\end{eqnarray}
where
\begin{eqnarray}
  H^{\rm 0PM}_0&=&-k_{t}^2+k_{r}^2+\frac{k_{\theta}^2}{r^2}
                  +\frac{k_{\varphi}^2}{r^2\sin^2\theta}, \\ 
  H^{\rm 1PM}_2&=&
  -\frac{2\tilde{M}}{r}k_{t}^2-\frac{2\tilde{M}}{r}k_r^2, \\
  H^{\rm 1.5PM}_3&=&
\frac{2D\tilde{a}}{1+D}\sqrt{\frac{2\tilde{M}}{r^5}}k_{r}k_{\varphi}, \\
  H^{\rm 2PM}_4&=&-\frac{4\tilde{M}^2}{r^2}k_{t}^2
+\frac{\tilde{a}^2\sin^2\theta}{(1+D)r^2}k_{r}^2
-\frac{\tilde{a}^2\cos^2\theta}{(1+D)r^4}k_{\theta}^2 \nonumber \\
&&\hspace{1.4cm}-\frac{\tilde{a}^2}{(1+D)r^4\sin^2\theta}k_{\varphi}^2
-\frac{4\tilde{a}\tilde{M}}{r^3}k_{t}k_{\varphi}, \\
 H^{\rm 3PM}_6&=&
-\frac{8(1+D)\tilde{M}^3-2\tilde{a}^2\tilde{M}\cos^2\theta}{(1+D)r^3}k_{t}^2 \nonumber \\
&&~~~~~+\frac{2\tilde{a}^2\tilde{M}(\cos^2\theta+D\sin^2\theta)}{(1+D)r^3}k_r^2 \nonumber \\
&&~~~~~~~-\frac{2(1+2D)\tilde{a}^2\tilde{M}}{(1+D)^2r^5}k_{\varphi}^2
-\frac{8\tilde{a}\tilde{M}^2}{r^4}k_{t}k_{\varphi}\,.
\end{eqnarray}
The superscripts showed the post-Minkowskian (PM) order.
$H^{\rm 0PM}_0$ produces a light-ray in the flat spacetime.
$H^{\rm 1PM}_2$ represents the lens effect due to the weak-field Schwarzschild potential.
The higher-order terms, $H^{\rm 2PM}_4$ and $H^{\rm 3PM}_6$, include 
the effects of the black hole spin and quadrupole moment.
For the photon motion, unlike the pulsar motion,
the leading correction due to the disformal parameter $D$
appears in $H^{\rm 1.5PM}_3$.
Since there is no corresponding term in the case of the Kerr solution,
this correction causes stronger light bending in the $\phi$-direction 
than in the usual Kerr correction.

\subsubsection{Motion of the S2-like pulsar and the initial conditions}
This work uses the Kepler motion to determine 
the initial conditions for the motion of the S2-like pulsar.
Let the coordinates of the orbital plane be $(X, Y)$. The Kepler motion with 
the eccentricity $e$ can be expressed by
\begin{eqnarray}
  X(u) &=& s(\cos{u}-e),\\
  Y(u) &=& s\sqrt{1-e^2}\sin{u},
\end{eqnarray}
where $u$ is the eccentric anomaly and $s$ is the semi-major axis.
The origin of $u$ is at the pericenter.
In the Kepler motion, the derivative of $u$
with respect to the coordinate time $t$ is given by
\begin{equation}
 \frac{du}{dt}=\sqrt{\frac{M_{\rm BH}}{s^3}}\frac{1}{1-e\cos u}\,.
\end{equation}
Thus, the components of the velocity on the orbital plane is given by
\begin{eqnarray}
  V_{X}(u)&=& \frac{dX}{dt}
  =-\sqrt{\frac{M_{\rm BH}}{s}}\frac{\sin{u}}{1-e\cos{u}}
  ,\\
  V_{Y}(u)&=&\frac{dY}{dt}
  = \sqrt{\frac{M_{\rm BH}}{s}}\frac{\sqrt{1-e^2}\cos{u}}{1-e\cos{u}}\,.
\end{eqnarray}
For the sake of simplicity, we fix a distant observer at $(x, y, z)=(0, 0, -R_0)$ 
in the pseudo-Cartesian coordinates.
Then, the position and the velocity of the Kepler motion
in the pseudo-Cartesian coordinates are given by
\begin{eqnarray}
  x(u) &=& (X\cos\omega-Y\sin\omega)\cos\Omega \nonumber \\
        &&\hspace{2cm} -(Y\cos\omega+X\sin\omega)\sin\Omega\cos{I}, \label{posi:x}\\
  y(u) &=& (Y\cos\omega+X\sin\omega)\cos\Omega\cos{I} \nonumber \\
        &&\hspace{2cm}+(X\cos\omega-Y\sin\omega)\sin\Omega,\\
  z(u) &=& (Y\cos\omega+X\sin\omega)\sin{I}, 
\end{eqnarray}
and
\begin{eqnarray}
  v_x(u) &=& (V_X\cos\omega-V_Y\sin\omega)\cos\Omega \nonumber \\
        &&\hspace{1.8cm}-(V_Y\cos\omega+V_X\sin\omega)\sin\Omega\cos{I},\\
  v_y(u) &=& (V_Y\cos\omega+V_X\sin\omega)\cos\Omega\cos{I}\nonumber \\
        &&\hspace{1.8cm}+(V_X\cos\omega-V_Y\sin\omega)\sin\Omega,\\
  v_z(u) &=&(V_Y\cos\omega+V_X\sin\omega)\sin{I}, \label{velo:z}
\end{eqnarray}
where $I$, $\omega$, and $\Omega$ are the 
inclination, the argument of periapsis, and the 
longitude of ascending node, respectively.
We choose the apocenter at $u=\pi$ as the initial position.
Finally, the initial conditions for the motion of the S2-like pulsar are given by
\begin{eqnarray}
  &&x_0=x(\pi),~~y_0=y(\pi),~~z_0=z(\pi),\\
  &&p_{x0}=v_x(\pi),~~p_{y0}=v_y(\pi),~~p_{z0}=v_{z}(\pi).
\end{eqnarray}
Given $x_0$, $y_0$, $z_0$, $p_{x0}$, $p_{y0}$, and $p_{z0}$, 
the Hamiltonian constraint that $H=-1/2$ becomes a quadratic equation for $p_{t}$
that is a conserved quantity of motion. We solve the quadratic equation and choose
the negative root.
To numerically solve Hamilton's equations for the pulsar, 
we use the Dormand--Prince method that is
the explicit fifth(forth)-order Runge--Kutta method implemented in {\it Mathematica}.
We should check if the Hamiltonian constraint is satisfied in the numerical integration.
The integration continues until the pulsar comes back to the apocenter again.
The relative integration errors are $\leq {\cal O}(10^{-15})$ during the calculation.

\subsubsection{Photon trajectories from the S2-like pulsar to the 
distant observer}
This study requires photon trajectories from the S2-like pulsar to the distant observer, 
but the initial $k_{\mu}$ that gives the trajectory is not known a priori.
Our procedure to find the initial $k_{\mu}$ is shown here.

We assume that a photon emitted at $\lambda=0$ hits 
the distant observer at $\lambda=1$.
Given the initial conditions for the photon motion,
the position of the photon at $\lambda=1$ is obtained 
by solving Hamilton's equations for the photon. 
Then, we can express the coordinate deviations between the photon position 
$x^{\rm hit}_i$ at $\lambda=1$ 
and the observer position $x^{\rm obs}_i$ as functions 
of the initial three-momentum of the photon $k^{\rm em}_j$ as follows
\citep{2010ApJ...711..157A}:
\begin{equation}
  f_{i}(k^{\rm em}_j(\tau);\tau)=x^{\rm hit}_i(k^{\rm em}_j(\tau))-x_i^{\rm obs},
\end{equation}
where the $\tau$ dependence of $k^{\rm em}_j$ has been explicitly shown with the argument, 
namely, $k^{\rm em}_j(\tau)=k_{j}|_{\lambda=0}$ is the three-momentum 
of an emitted photon from the pulsar at $(x(\tau), y(\tau), z(\tau))$.
This work uses the Newton--Raphson method to 
find  $k^{\rm em}_j$ that satisfies $f_i(k^{\rm em}_j)=0$.
In practice, we give the criterion for $k_j^{\rm em}$ as
\begin{eqnarray}
  &&\sqrt{(x^{\rm hit}-x^{\rm obs})^2+(y^{\rm hit}-y^{\rm obs})^2
+(z^{\rm hit}-z^{\rm obs})^2} \nonumber \\
  &&\hspace{5.6cm}<10^{-7}M_{\rm BH}.
\end{eqnarray}
Because the radius of a neutron star would be $\sim 10^{-5}M_{\rm BH}$, 
the above criterion is sufficient for our purpose.
Moreover, we set the distant observer at 
$(x^{\rm obs}, y^{\rm obs}, z^{\rm obs})=(0, 0, -10^{8}M_{\rm BH})$ 
rather than $(0, 0, -R_0)$, where $R_0\sim 10^{10}M_{\rm BH}$, 
to search $k_j^{\rm em}$ efficiently \citep{2015ApJ...809..127Z, 2017ApJ...849...33Z}.
The changes of the observables due to the replacement of the observer position 
are smaller than the observational uncertainties in the SKA.
To obtain $x_i^{\rm hit}$, we numerically solve Hamilton's equations for the photon
with the Dormand--Prince method.
 $k^{\rm em}_{t}$ is determined by the Hamiltonian constraint that $H=0$ 
with the initial conditions of the photon.
$k^{\rm em}_{t}$ is the conserved quantity of motion, and then 
we check whether the Hamiltonian constraint is satisfied during the calculation.
We see that $|H|/(k^{\rm em}_{t})^2\leq {\cal O}(10^{-14})$ 
in the integration.

\subsection{Apparent position and the time of arrival}
We show our formalism to express the position of the S2-like pulsar on the sky
and the arrival time of the pulse here.
This study focuses on the gravitational effect due to the disformal Kerr black hole.
We ignore other environmental effects such as the effect of the interstellar medium. 
\subsubsection{Apparent position on the sky}
The pulsar's position is given on the sky plane $(\alpha, \beta)$, 
where $\alpha$ is the right accession and $\beta$ is the declination angle.
We can relate the angular position to the apparent position 
in the pseudo-Cartesian coordinates, $(x_{\rm ap}, y_{\rm ap})$, as
\begin{equation}
  \alpha=-\frac{y_{\rm ap}}{R_0},~~\beta=\frac{x_{\rm ap}}{R_0}.
  \label{eq:ap_posi}
\end{equation}
The apparent position $(x_{\rm ap}, y_{\rm ap})$ can be 
determined by the four-momentum of an emitted photon arriving at the distant observer.
Let the four-momentum of the photon at $\lambda=1$ be $k^{\rm hit}_{\mu}$, that is,
$k^{\rm hit}_{\mu}=k_{\mu}|_{\lambda=1}$.
From the view of the observer, the photon is coming from the direction
of the vector $k^{\rm hit}_{\mu}$.
Thus, solving the equation of motion for the photon with $-k^{\rm hit}_{\mu}$ in 
the Minkowski spacetime, we can obtain the apparent position of the pulsar 
in the pseudo-Cartesian coordinates as follows:
\begin{equation}
  x_{\rm ap} = -k_{x}^{\rm hit},~~  y_{\rm ap} = -k_{y}^{\rm hit}.
 \label{eq:ap_posi2}
\end{equation}
Usually, the proper motion of Sgr A* on the sky is added to equation (\ref{eq:ap_posi})
[e.g., see Zhang and Saha (\yearcite{2017ApJ...849...33Z}) for pulsar studies].
Because the proper motion of Sgr A* is canceled out when we see the difference between 
the disformal Kerr and Kerr, this study uses equation (\ref{eq:ap_posi}) with equation 
(\ref{eq:ap_posi2}) for the angular position of the S2-like pulsar.

\subsubsection{Time of arrival}
We calculate the time of arrival (TOA) of the pulse without environment 
effects between the central black hole and the distant observer for the sake of simplicity.
Then, the TOA is purely determined by the redshift of emitted photons, $Z$, 
which can be written by
\begin{equation}
  Z=\frac{p^{\mu}k_{\mu}^{\rm em}}{U^{\mu}_{\rm obs}k^{\rm hit}_{\mu}}-1,
  \label{eq:redshift}
\end{equation}
where $U^{\mu}_{\rm obs}$ is the four-velocity of the distant observer 
given by $U_{\rm obs}^{\mu}=(1, 0, 0, 0)$ in the pseudo-Cartesian coordinates.
We easily see that $U^{\mu}_{\rm obs}k_{\mu}^{\rm hit}=k^{\rm hit}_{t}=k^{\rm em}_{t}$ 
because the time component of $k_{\mu}$ is the conserved quantity of motion.
Thus, equation (\ref{eq:redshift}) can be expressed as 
\begin{equation}
  Z=\frac{p^{\mu}k_{\mu}^{\rm em}}{k^{\rm em}_{t}}-1.
 \end{equation}
$p_{\mu}$ and $k_{\mu}^{\rm em}$ are given at each proper time of the S2-like pulsar.
Then, we can calculate the TOA by the following integral with respect to the proper time 
\citep{2017ApJ...849...33Z}:
\begin{equation}
  t_{\rm TOA}(\tau)=\int^{\tau}_0(Z(\tilde{\tau})+1)d\tilde{\tau}.
  \label{eq:TOA}
\end{equation}
It would be helpful to show the approximated 
redshift in the post-Newtonian and 
post-Minkowskian approximations.
The redshift can be expanded up to 
$2$PN+$3$PM order (the order of $\epsilon^6$)  as follows:
\begin{eqnarray}
 &&Z+1 \nonumber \\
&&\approx  -p_t+\epsilon\left(\frac{p_rk_r^{\rm em}}{k_t^{\rm em}}
       +\frac{1}{r^2}\frac{p_\theta k^{\rm em}_\theta}{k^{\rm em}_t}
       +\frac{1}{r^2\sin^2\theta}\frac{p_\varphi k_\varphi^{\rm em}}{k_t^{\rm em}}\right) \nonumber \\
&&~~~~~-\epsilon^2\frac{2\tilde{M}}{r}p_t
-\epsilon^3\frac{2\tilde{M}}{r}\frac{p_rk_r^{\rm em}}{k_t^{\rm em}}\nonumber \\
&&~~~~~+\epsilon^4\Biggl{(}-\frac{4\tilde{M}^2}{r^2}p_t
-\frac{2\tilde{a}\tilde{M}}{r^3}\frac{p_tk_\varphi^{\rm em}}{k_t^{\rm em}} \nonumber \\
&&\hspace{1.6cm}+\frac{D\tilde{a}}{1+D}\sqrt{\frac{2\tilde{M}}{r^5}}
\frac{p_{r}k^{\rm em}_{\varphi}}{k^{\rm em}_{t}}
     +\frac{D\tilde{a}}{1+D}\sqrt{\frac{2\tilde{M}}{r^5}}
\frac{p_{\varphi}k_{r}^{\rm em}}{k^{\rm em}_{t}} \Biggr{)} \nonumber \\
&&~~~~~+\epsilon^5\Biggl{(}
-\frac{2\tilde{a}\tilde{M}}{r^3}p_\varphi
+\frac{\tilde{a}^2\sin^2\theta}{(1+D)r^2}\frac{p_rk_r^{\rm em}}{k_t^{\rm em}} \nonumber\\
&&\hspace{1.6cm}
-\frac{\tilde{a}^2\cos^2\theta}{(1+D)r^4}\frac{p_\theta k_{\theta}^{\rm em}}{k_t^{\rm em}}
-\frac{\tilde{a}^2}{(1+D)r^4\sin^2\theta}\frac{p_\varphi k_{\varphi}^{\rm em}}{k_t^{\rm em}}
\Biggr{)} \nonumber \\
&&~~~~~+\epsilon^6\left(
-\frac{8(1+D)\tilde{M}^3-2\tilde{a}^2\tilde{M}\cos^2\theta}{(1+D)r^3}p_t
-\frac{4\tilde{a}\tilde{M}^2}{r^4}\frac{p_t k_\varphi^{\rm em}}{k_t^{\rm em}}
\right)\,.\nonumber \\
\end{eqnarray}
We find that the leading correction of the TOA due to the disformal parameter $D$
appears in the $1.5$PN+$2$PM order (the order of $\epsilon^4$).

\subsection{Test case: differences between the Kerr and the Schwarzschild}
\begin{figure*}[h]
\begin{center}
\subfigure[]
{
\includegraphics[width=70mm]{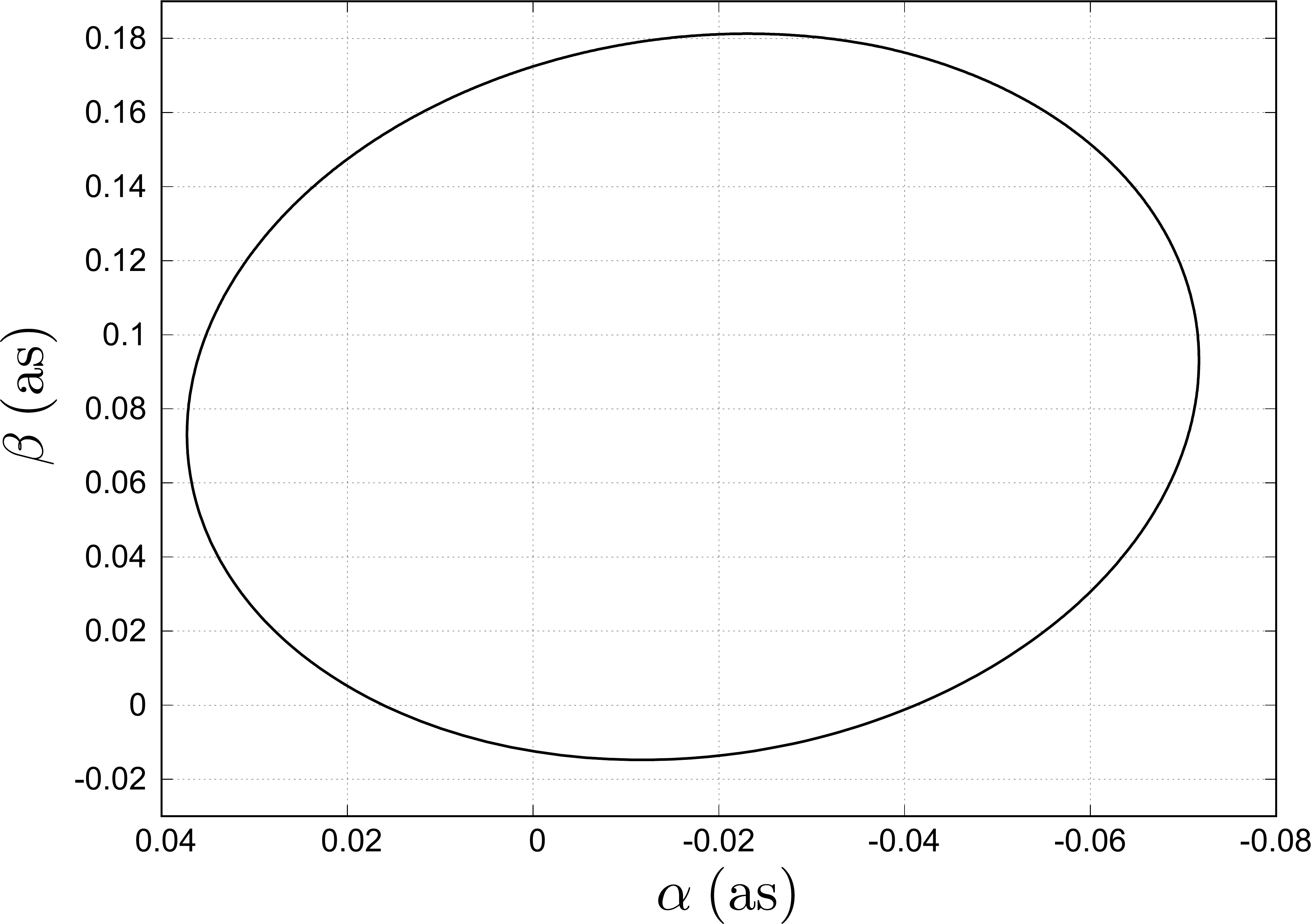}
}
\subfigure[]
{
\includegraphics[width=70mm]{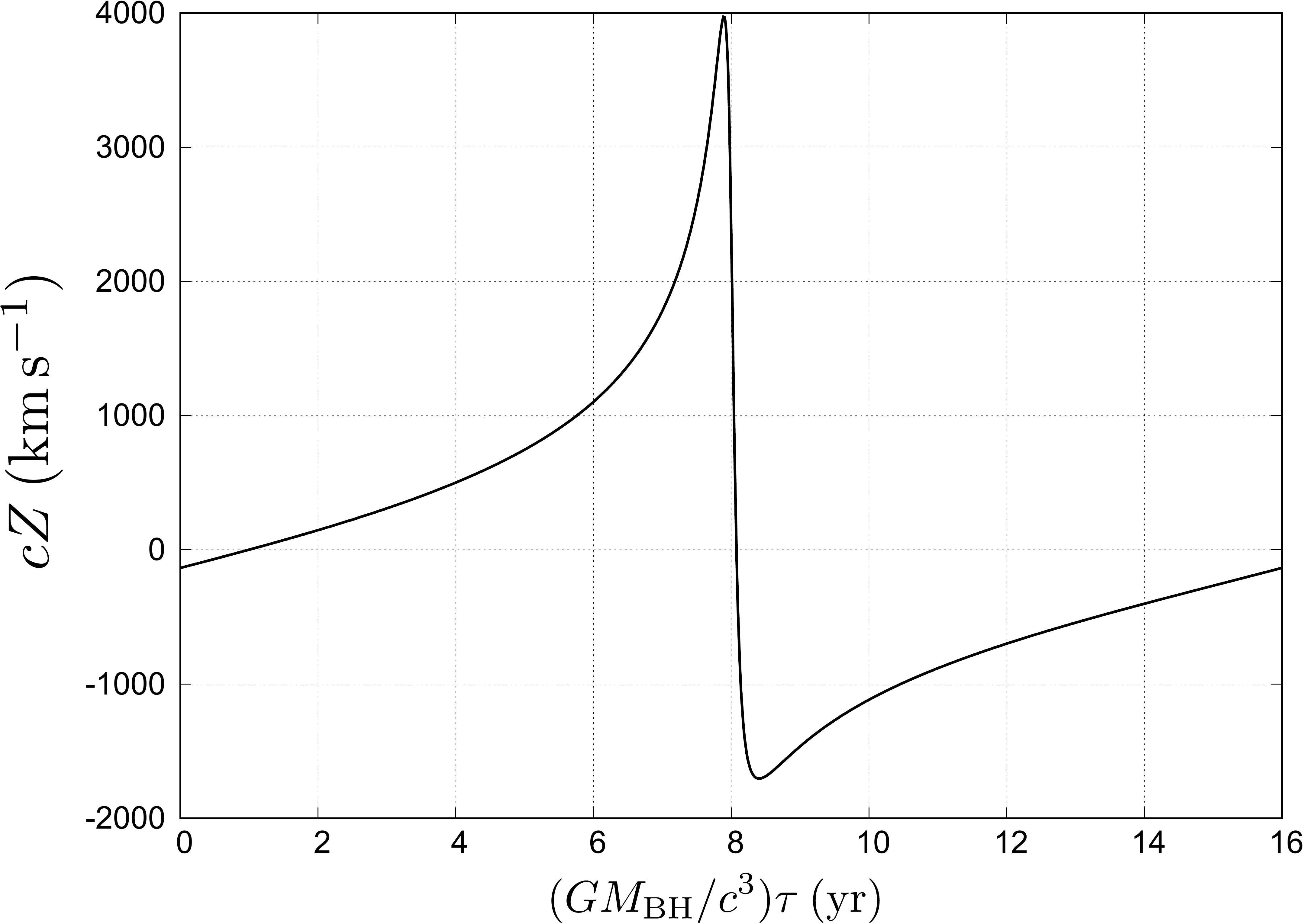}
}
\end{center}
\caption{Apparent position of the S2-like pulsar (a) 
and time evolution of the redshift of photons from the pulsar (b).
The time used in the panel (b) is the proper time of the S2-like pulsar.}
\label{fig:S2}
\end{figure*}

\begin{figure*}[h]
\begin{center}
\subfigure[]
{
\includegraphics[width=70mm]{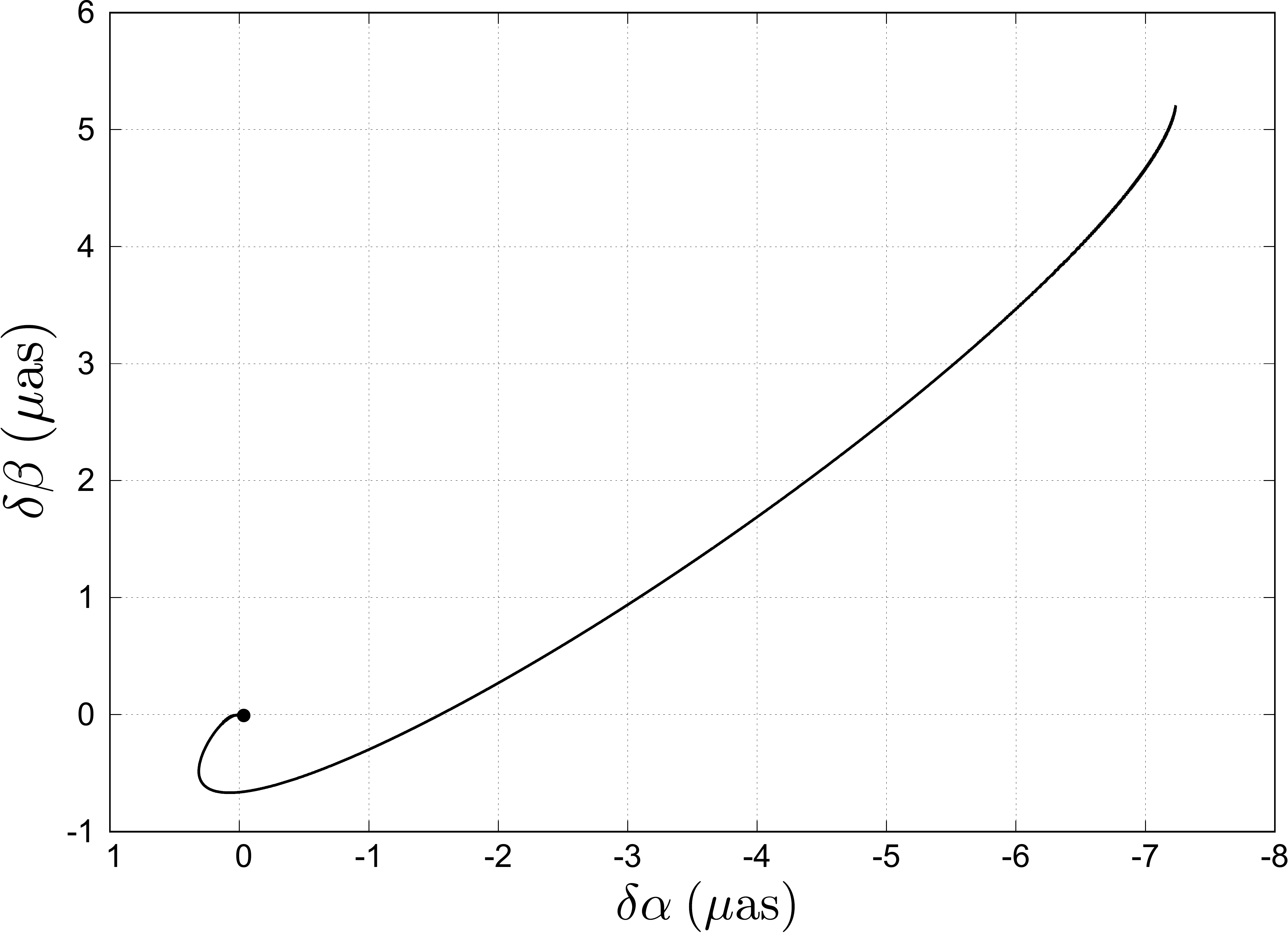}
\label{fig:devKerr_Schw_left}
}
\subfigure[]
{
\includegraphics[width=70mm]{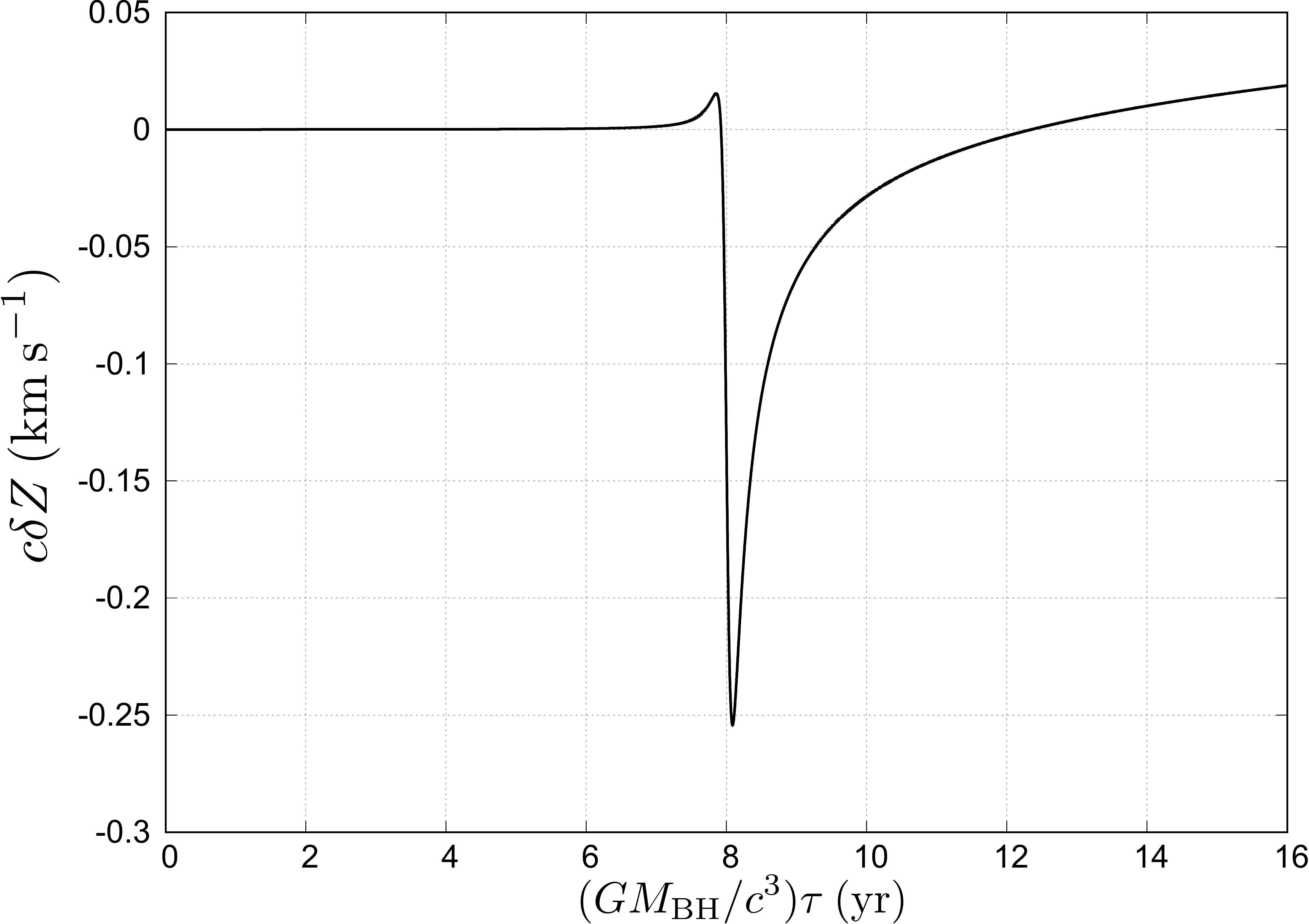}
\label{fig:devKerr_Schw_right}
}
\end{center}
\caption{Differences between the Kerr and the Schwarzschild.
The difference in the apparent position, $(\delta\alpha, \delta\beta)$, is in the panel (a). 
The filled circle represents the difference at $\tau=0$, and 
the difference does not return to the initial position due to the apocenter shift.
The difference in the redshift, $c\delta Z$, 
is in the panel (b). As in the panel (a), we can see the secular effect due to 
the apocenter shift.}
\label{fig:devKerr_Schw}
\end{figure*}

\begin{figure}[h]
  \begin{center}
  \includegraphics[width=70mm]{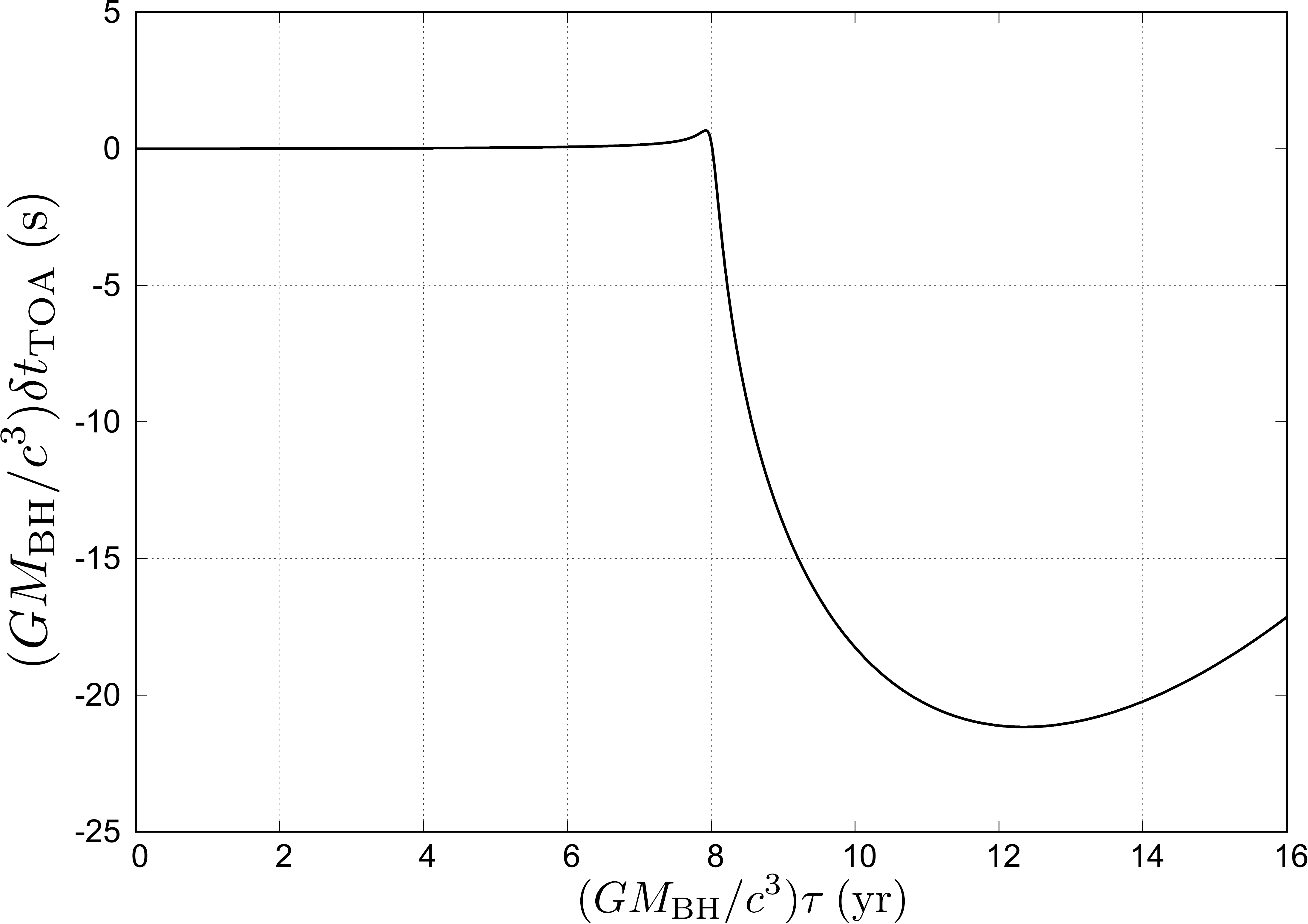}    
  \end{center}
  \caption{Spin-induced effect in $t_{\rm TOA}$.
    $\delta t_{\rm TOA}$ represents the difference in the 
    TOA between the Kerr and the Schwarzschild.}
  \label{fig:devToA_KerrSchw}
\end{figure}

We show the Kerr case as a test before investigating the 
disformal Kerr black hole with various sets of $(\tilde{a}, D)$.
We calculate the apparent position and the TOA for the S2-like pulsar 
in the extremal Kerr $(\tilde{a}_{\ast}, D)=(1, 0)$ and 
the Schwarzschild $(\tilde{a}_{\ast}, D)=(0, 0)$ cases, where 
$\tilde{a}_\ast=\tilde{a}/\tilde{M}$. 
In figure {\ref{fig:S2}}, we show the apparent position and the time evolution
of the redshift measured in the proper time $\tau$ in the case of 
$(\tilde{a}_{\ast}, D)=(1, 0)$. 
Then, we compare the results of those cases and show the differences in figure \ref{fig:devKerr_Schw}.
Here, we express the difference between the values in the cases of the extremal Kerr 
and the Schwarzschild by $\delta$, namely, for example, 
$\delta\alpha=\alpha|_{\tilde{a}_\ast=1}-\alpha|_{\tilde{a}_\ast=0}$.
The spin-induced difference in the apparent position is on the order of $1\>\mu{\rm as}$, and, 
for the redshift, the maximum of the difference is on the order of 
$10^{-1}\>{\rm km\cdot s^{-1}}$.
These values are consistent with Zhang, Lu, and Yu (\yearcite{2015ApJ...809..127Z}).
The spin-induced difference in the TOA is shown in figure \ref{fig:devToA_KerrSchw}, 
which is consistent with Zhang and Saha (\yearcite{2017ApJ...849...33Z}). 
The expected astrometric accuracy of the SKA is $10\>\mu{\rm as}$.
In figure \ref{fig:devKerr_Schw_left}, the spin-induced effects in the apparent position  
are smaller than the astrometric accuracy $10\>\mu{\rm as}$ within 
the time interval taken for the figure. 
However, the difference secularly increases as the S2-like pulsar orbits Sgr A*.
Then the difference in the apparent position becomes detectable  within 
a few periods $(<50\>{\rm yr})$.
For the TOA, the accuracy of the SKA is from $0.1\>{\rm ms}$ to $10\>{\rm ms}$, 
which varies with the observed frequencies.
From figure \ref{fig:devToA_KerrSchw},
the spin-induced effect is obviously far larger than its accuracy, 
and it is detectable with the SKA \citep{2017ApJ...849...33Z}.


\section{Effects of the disformal parameter $D$ and
its detectability with the SKA}
\label{sec:results}
We investigate the disformal Kerr black hole with various sets of $(\tilde{a}_\ast, D)$
and show the effects of the disformal parameter $D$.
Moreover, we discuss the detectability of the disformal parameter $D$ with the SKA.
The parameters in our simulations are listed in table \ref{tab:study_cases}.
In Case I, the spin parameter is set to be $\tilde{a}_{\ast}=0.1, 0.5$, and $0.7$ 
and we examine the motion of the S2-like pulsar.
The parameter range of $D$ in which the disformal Kerr spacetime describes a 
black hole spacetime depends on the value of $\tilde{a}_\ast$.
Let $D_{+}(\tilde{a}_\ast)$ and $D_{-}(\tilde{a}_\ast)$ denote
the upper and lower values of $D$ for a fixed value of $\tilde{a}_\ast$, respectively.
For $\tilde{a}_\ast\leq0.5$, there is no upper value for $D$, that is, 
we can take $D_{+}\rightarrow\infty$. 
Alternatively, the lower value, $D_{-}$, is obtained.
From \citet{2021JHEP...01..018A}, we obtain $D_{-}$ as 
$D_{-}\sim-0.96$ for $\tilde{a}_\ast=0.1$, and $D_{-}\sim-0.41$ for $\tilde{a}_\ast=0.5$.
In the case $\tilde{a}_\ast>0.5$, both the upper and lower values are obtained.
For $\tilde{a}_\ast=0.7$, we obtain $D_{-}\sim-0.17$ and $D_{+}\sim0.53$.
We pick up some $D$ within $D_{-}<D<D_{+}$ for each $\tilde{a}_\ast$.

In Case II and III, setting $(\tilde{a}_\ast, D)=(0.5, 1)$, 
we change the inclination angle to $I=180^{\circ}$~(face-on) and 
$I=95^{\circ}$~(nearly edge-on) for the S2-like pulsar, respectively, and
consider three values of the eccentricity: $e=0$, $0.4$, and $0.88$.
Note that when using the inclination angle that $I=90^{\circ}$~(edge-on), 
pulsars may almost overlap behind the central black hole on the line-of-sight. 
Then we have to seriously take the strong gravitational lensing effect due to 
the black hole into account. 
To avoid this situation, this study uses the value that $I=95^{\circ}$.

\begin{table*}[htbp]
  \caption{Summary of parameter sets for Case I~(S2-like), 
      Case II~(face-on) and III~(nearly edge-on). } \label{tab:study_cases}
   \vspace{0.1cm}
  \begin{tabular}{cccc}\hline
    Case~&~~$\tilde{a}_{\ast}=\tilde{a}/M_{\rm BH}$~~&~~$D$~~
    &~$(e, T, I, \omega, \Omega)$\\
    \hline
    I(a) & 0.1 &~-0.9, -0.8, -0.6, 1, 10, and 1000~&~~$(0.88, 16, 135, 65, 225)$\\
    I(b) & 0.5 &~-0.4, -0.2, 0.5, 1, 10, and 1000~&~~$(0.88, 16, 135, 65, 225)$\\
    I(c) & 0.7 &~-0.1, 0.1, 0.2, and 0.5~&~~$(0.88, 16, 135, 65, 225)$\\
    \hline
    II(a) & 0.5 & 1 &~~$(0, 16, 180, 65, 225)$\\
    II(b) & 0.5 & 1 &~~$(0.4, 16, 180, 65, 225)$\\
    II(c) & 0.5 & 1 &~~$(0.88, 16, 180, 65, 225)$\\
    \hline
    III(a) & 0.5 & 1 &~~$(0, 16, 95, 65, 225)$\\
    III(b) & 0.5 & 1 &~~$(0.4, 16, 95, 65, 225)$\\
    III(c) & 0.5 & 1 &~~$(0.88, 16, 95, 65, 225)$\\
    \hline
  \end{tabular}
\end{table*}

\subsection{Results}
For each case in table \ref{tab:study_cases}, the apparent position and the TOA are obtained.
Moreover, we perform our simulations without the disformal parameter $D$ and then compare
the results with and without $D$. 
The difference is denoted by $\delta$ as in the previous section, namely, 
$\delta\alpha=\alpha|_{(\tilde{a}_{\ast}=0.1, D=1)}-\alpha|_{(\tilde{a}_{\ast}=0.1, D=0)}$, 
for example. 
We demonstrate the results for Case I, II, and III in the following subsections, respectively.

\subsubsection{Effects of the disformal parameter $D$ at 
$1.5$PN, $2$PN and higher PN order terms}
\begin{figure*}[htbp]
\begin{center}
\subfigure[]
{
\includegraphics[width=70mm]{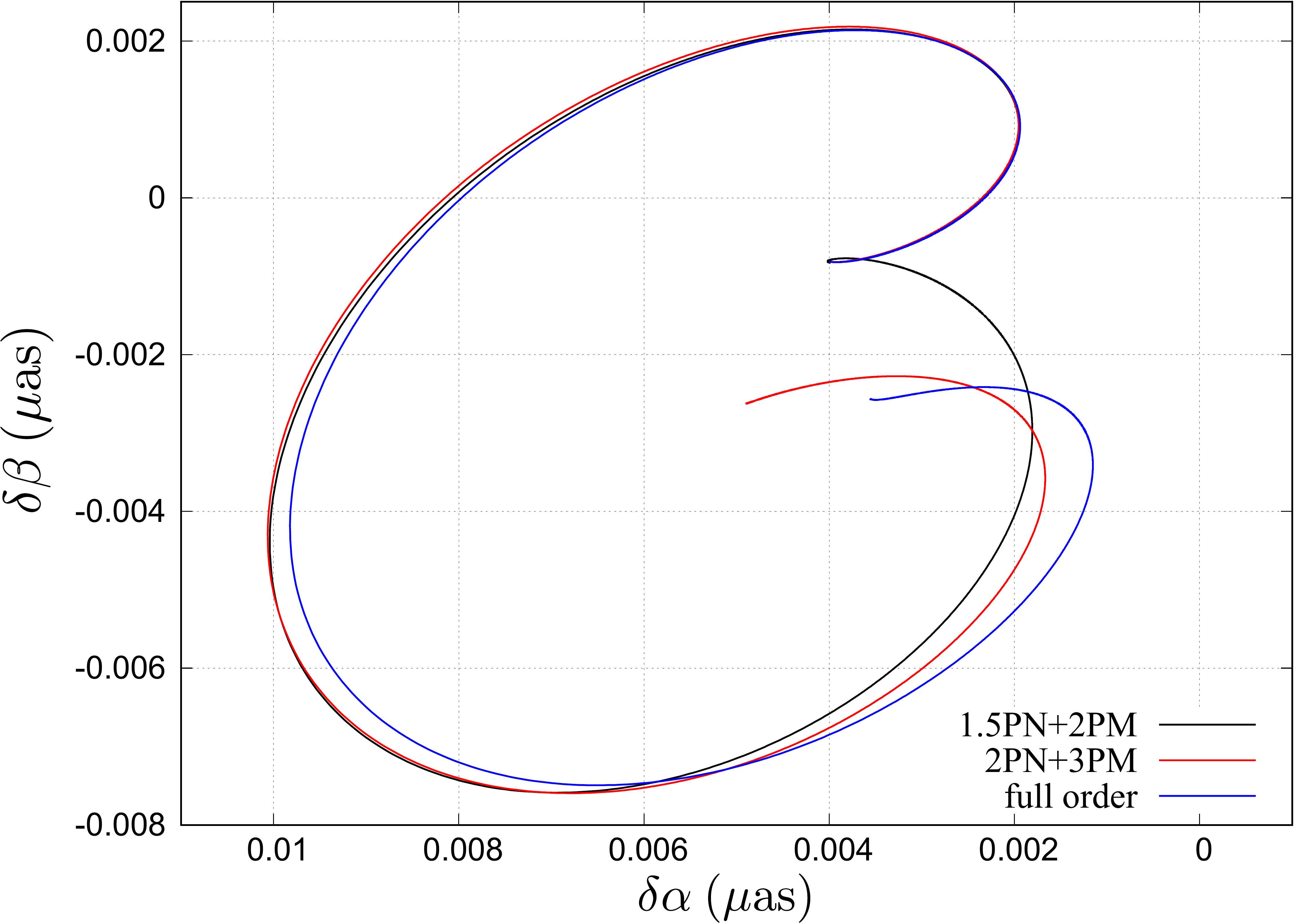}
\label{fig:devDKerr_Kerr_15PN2PN_left}
}
\subfigure[]
{
\includegraphics[width=70mm]{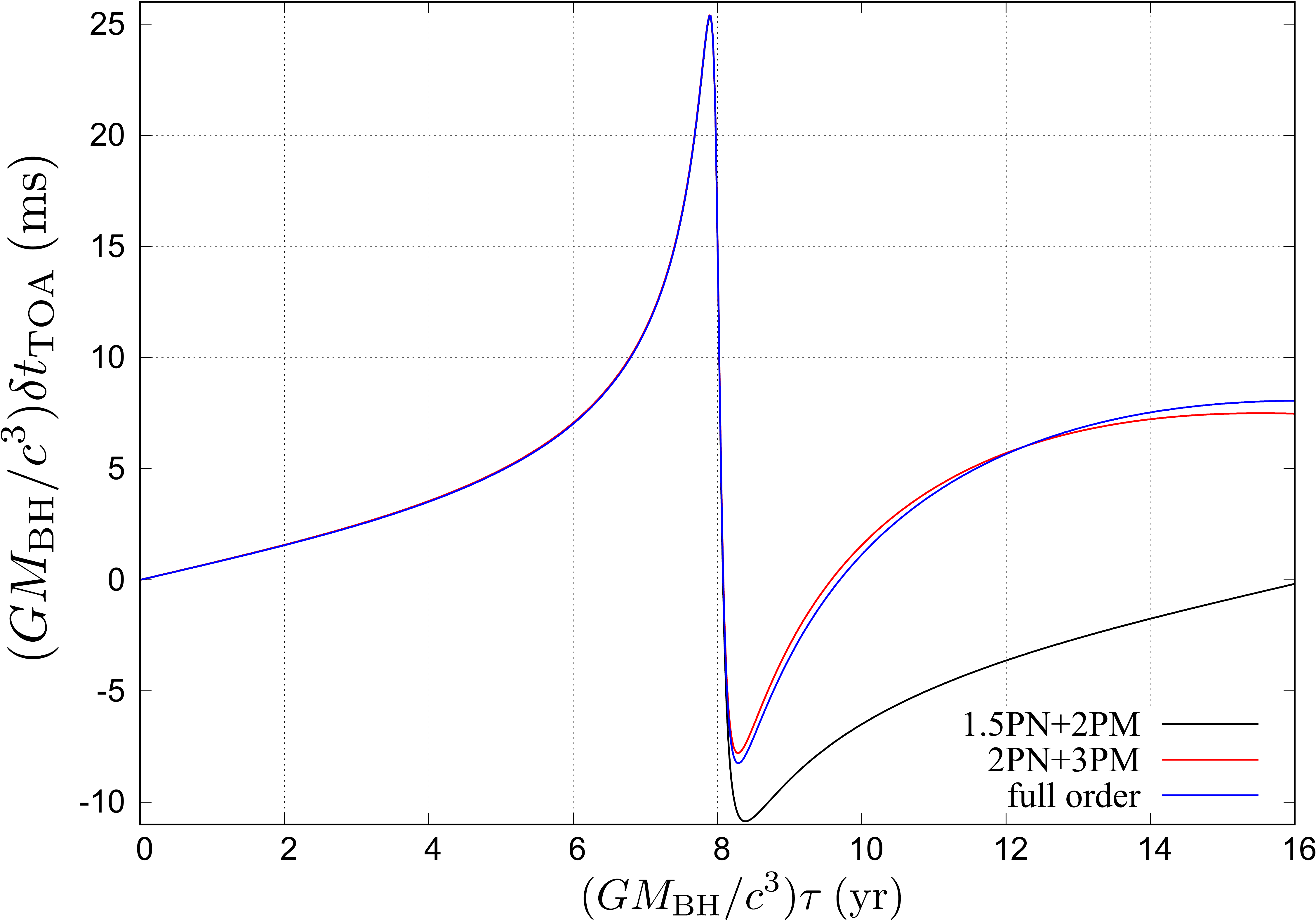}
\label{fig:devDKerr_Kerr_15PN2PN_right}
}
\end{center}
\caption{
Differences between the cases that $(\tilde{a}_\ast, D)=(0.5, 1)$ and $(0.5, 0)$.
The panel (a) shows the difference in the apparent position and (b) is that in the TOA.
In the panels, the black and red solid lines represent the results from 
the $1.5$PN+$2$PM and $2$PN+$3$PM approximations,
respectively. The blue solid lines come from the full order equations.
We see that a significant apocenter shift appears 
in the $2$PN+$3$PM approximation from both the panels.
We also find that the higher-order terms in the full order equations 
do not add any significant corrections to the results 
in the $2$PN+$3$PM approximation.
In the panel (a), $(\delta\alpha, \delta\beta)$ 
at $\tau=0$ is off the origin.
Although the initial positions of the S2-like pulsar are the same in both
the disformal Kerr and Kerr cases, 
the apparent positions are different from each other 
in general because of the difference in the photon trajectory. (Color online)}
\label{fig:devDKerr_Kerr_15PN2PN}
\end{figure*}

The effect of the disformal parameter $D$ appears from $1.5$PN order 
in the Hamiltonian for the S2-like pulsar. 
To see the significance of the terms at each order, 
we calculate the apparent position and the TOA 
taking the terms up to 1.5PN+2PM order (the order of $\epsilon^5$ for the pulsar motion
and $\epsilon^4$ for photon motion and the redshift), 
2PN+3PM order (the order of $\epsilon^6$),
and full order (without the PN and PM expansions)
and compare them with each other.
We choose Case I(b) where $(\tilde{a}_\ast, D)=(0.5, 1)$ and calculate 
$\delta\alpha$, $\delta\beta$, and $\delta t_{\rm TOA}$ in each order.
The differences in each order are shown in figure \ref{fig:devDKerr_Kerr_15PN2PN}.
From the figures, we find that the deviation from the Kerr
is mostly determined by the $1.5$PN+$2$PM approximation.
It means that the non-circularity of the spacetime
at the $1.5$PN order, which also gives the $1.5$PM order,
mainly produces the deviation.
From figure \ref{fig:devDKerr_Kerr_15PN2PN_left},  
we can see the apocenter shift due to the disformal parameter $D$ 
appears in the case of the $2$PN+$3$PM approximation.
Although the effect of the disformal parameter $D$
appears at the $1.5$PN order in the pulsar motion, 
a significant secular apocenter shift appears at the $2$PN order.
It is consistent with Anson,  Babichev, and Charmousis (\yearcite{2021PhRvD.103l4035A}) 
where they have shown that the secular shift of the motion appears at 2PN order in generic $D$.
We can also see the effect of the apocenter shift in 
$\delta t_{\rm TOA}$ from figure~\ref{fig:devDKerr_Kerr_15PN2PN_right}.
The TOA is given by the integration of the redshift, and therefore
the secular shift due to the $2$PN order becomes significant in 
the second half of the orbit.
Since we do not find any significant corrections from the higher-order 
terms in the post-Newtonian and post-Minkowskian approximations 
in figure \ref{fig:devDKerr_Kerr_15PN2PN}, hereafter, we 
show the results calculated using equations up to $2$PN+$3$PM order.

\subsubsection{Case I~(S2-like: inclination angle $I=135^\circ$)}
\begin{figure*}[htbp]
\begin{center}
\subfigure[]
{
\includegraphics[width=70mm]{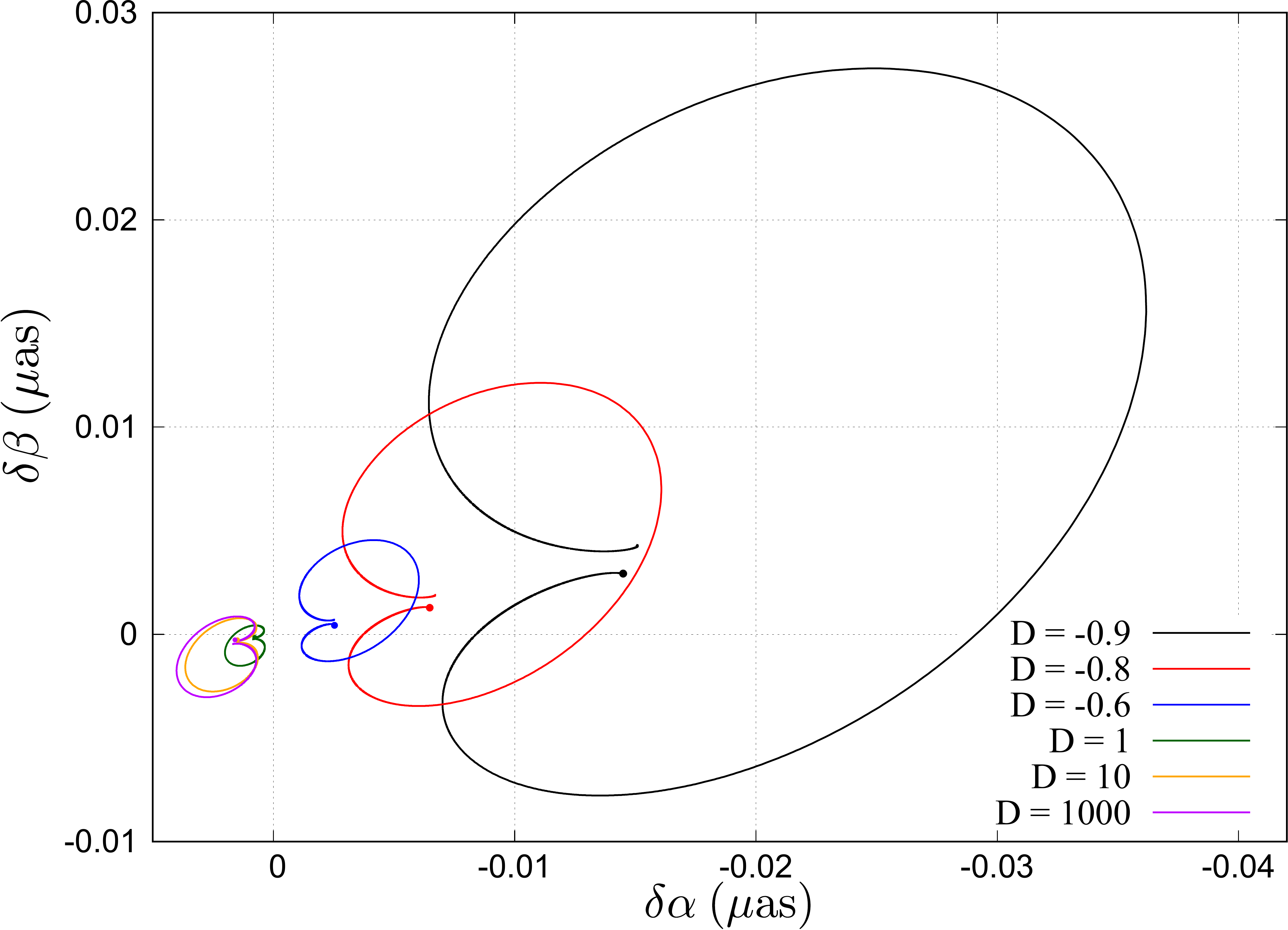}
\label{fig:devDKerr_Kerr_a01_left}
}
\subfigure[]
{
\includegraphics[width=70mm]{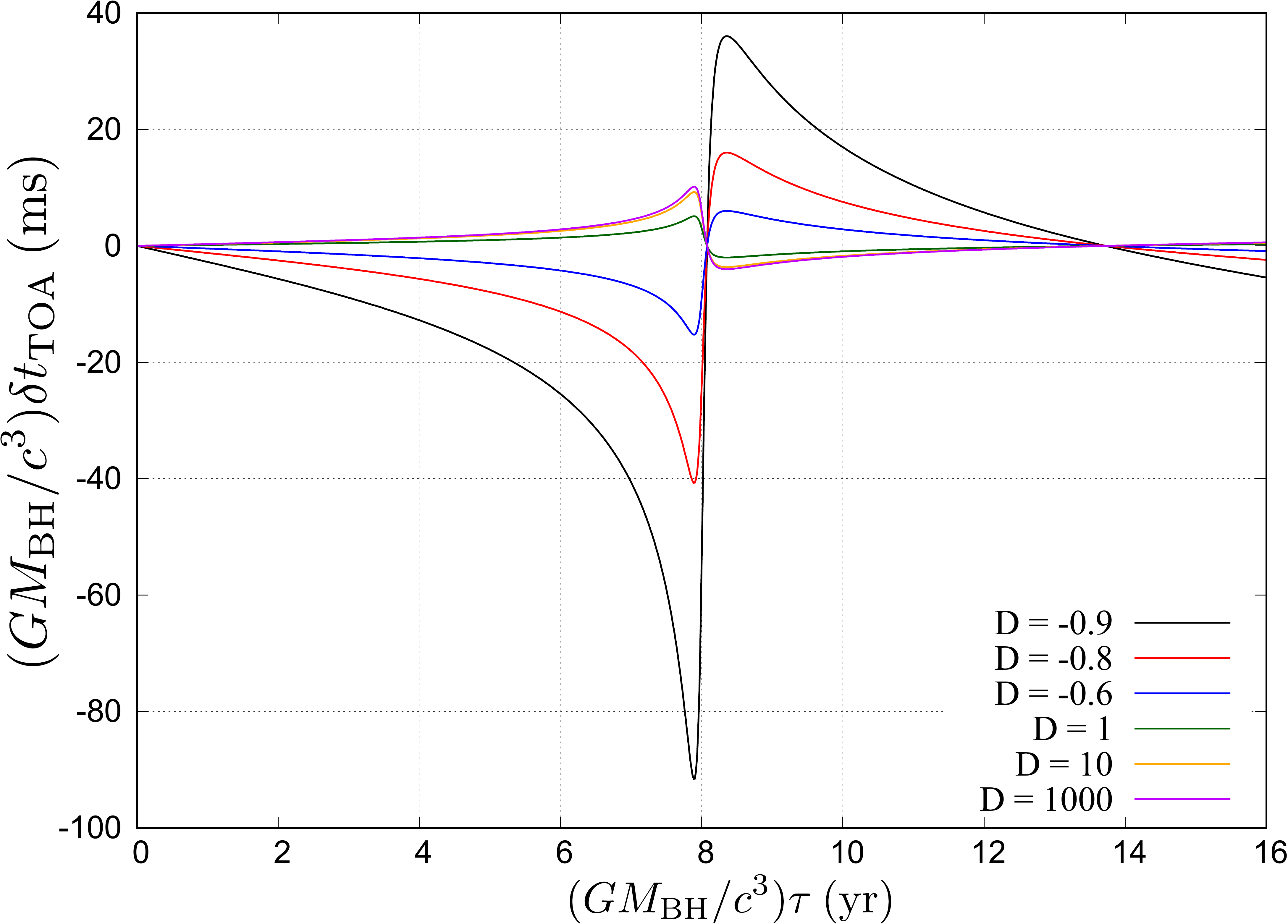}
\label{fig:devDKerr_Kerr_a01_right}
}
\end{center}
\caption{Differences in the apparent position (a) and the TOA (b) 
in the case $\tilde{a}_\ast=0.1$.
We show $(\delta\alpha, \delta\beta)$ and $\delta t_{\rm TOA}$ in the cases 
that $D=-0.9\,({\rm black}), -0.8\,({\rm red}), -0.6\,({\rm blue}), 1\,({\rm green}), 
10\,({\rm orange})$, and $1000\,({\rm magenta})$.
The filled circles in the panel (a) represent $(\delta\alpha, \delta\beta)$ at 
$\tau=0$. (Color online)}
\label{fig:devDKerr_Kerr_a01}
\end{figure*}

\begin{figure*}[htbp]
\begin{center}
\subfigure[]
{
\includegraphics[width=70mm]{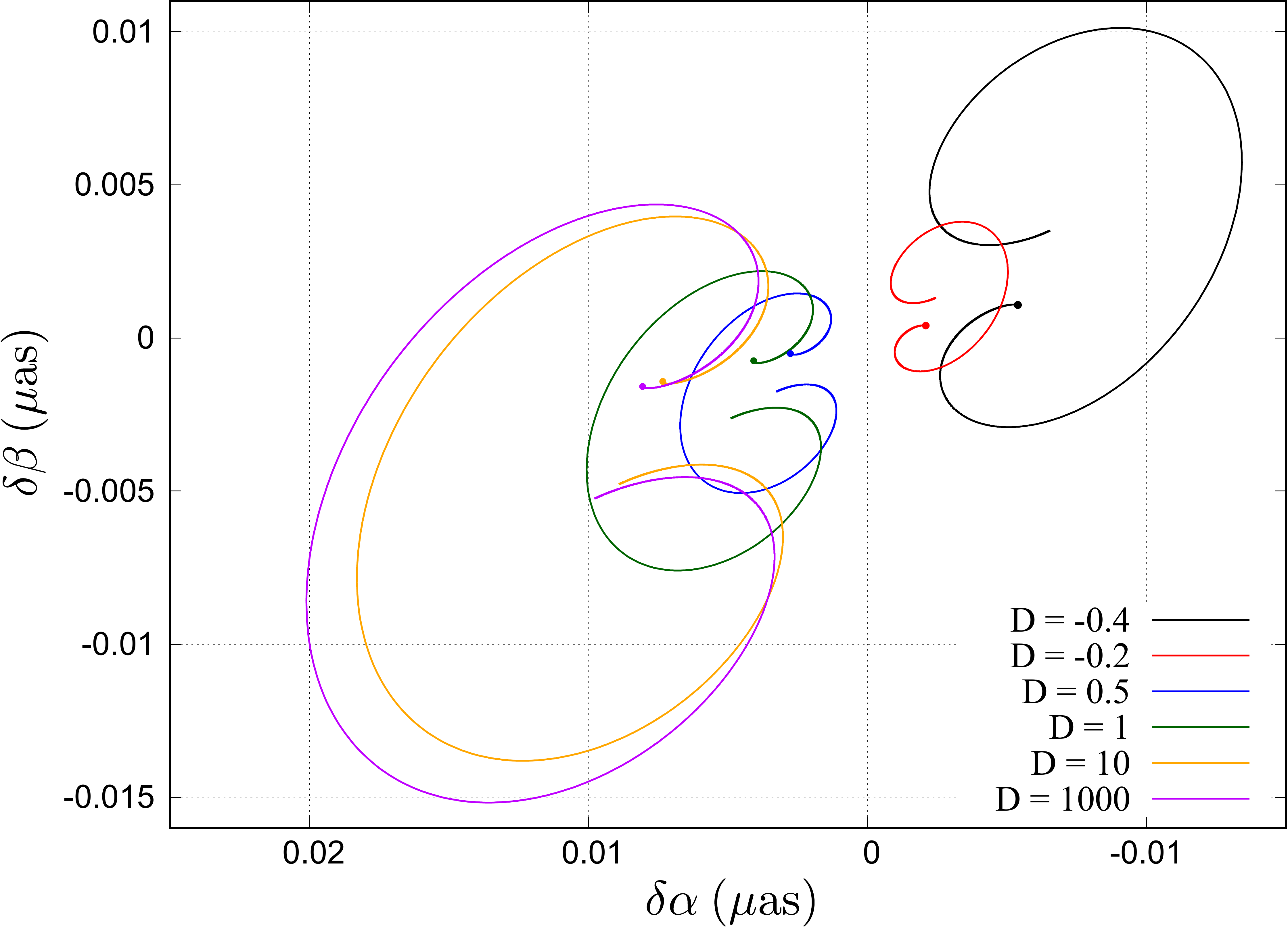}
\label{fig:devDKerr_Kerr_a05_left}
}
\subfigure[]
{
\includegraphics[width=70mm]{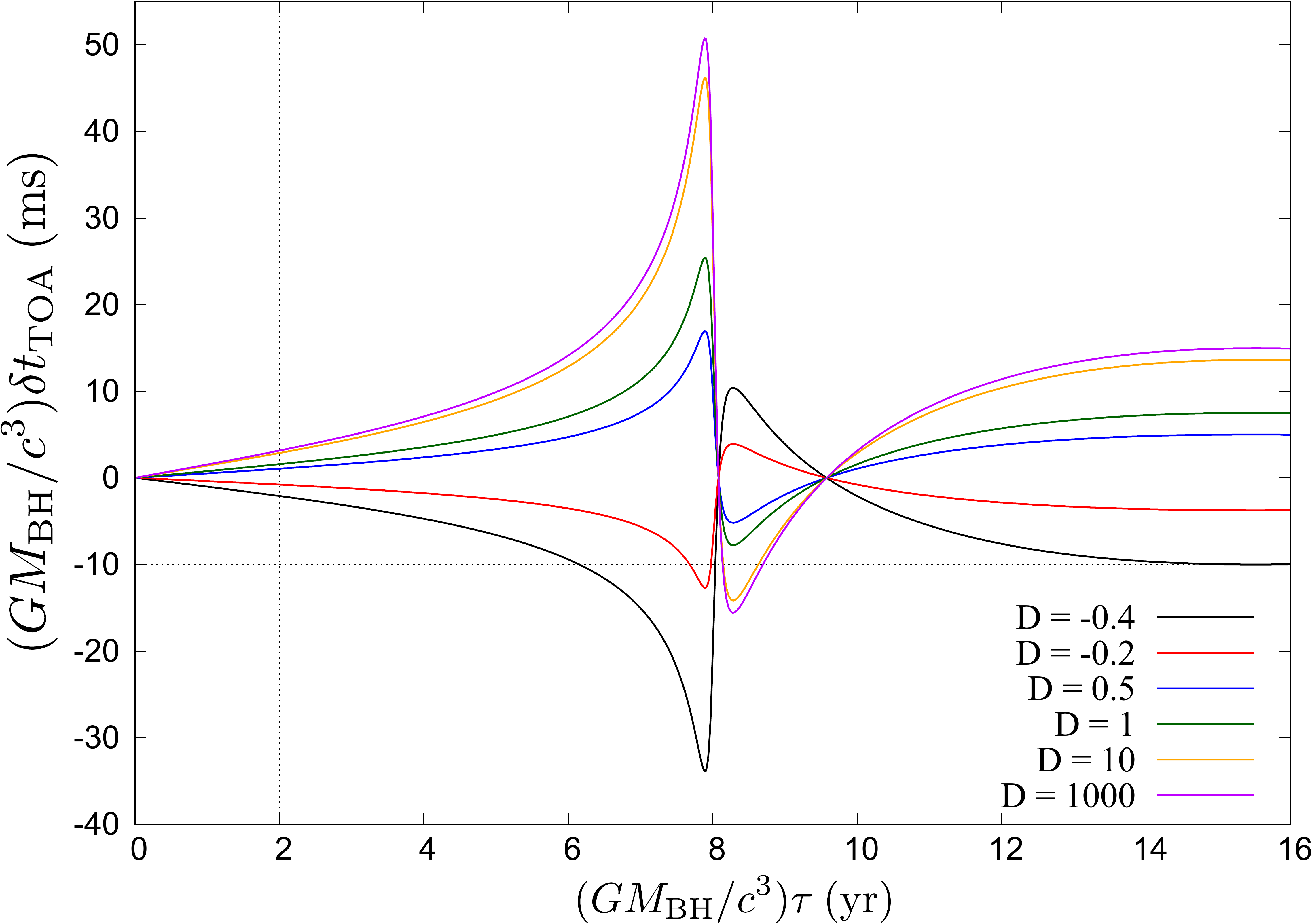}
\label{fig:devDKerr_Kerr_a05_right}
}
\end{center}
\caption{Differences in the apparent position (a) and the TOA (b) in the case 
$\tilde{a}_\ast=0.5$. 
The solid lines with colors represent the cases that
$D=-0.4\,({\rm black}), -0.2\,({\rm red}), 0.5\,({\rm blue}), 
1\,({\rm green}), 10\,({\rm orange})$, and $1000\,({\rm magenta})$. (Color online)}
\label{fig:devDKerr_Kerr_a05}
\end{figure*}

\begin{figure*}[htbp]
\begin{center}
\subfigure[]
{
\includegraphics[width=70mm]{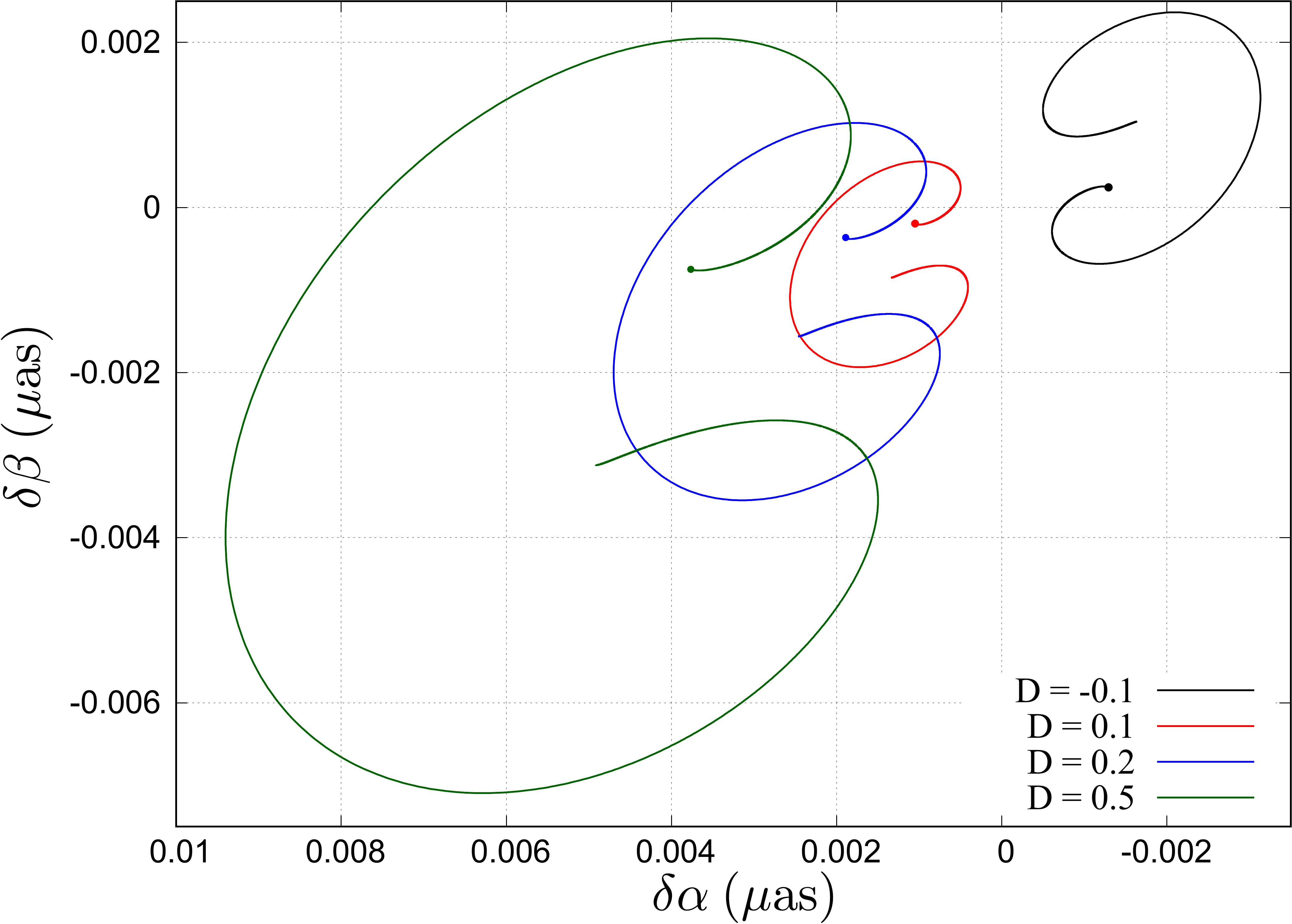}
\label{fig:devDKerr_Kerr_a07_left}
}
\subfigure[]
{
\includegraphics[width=70mm]{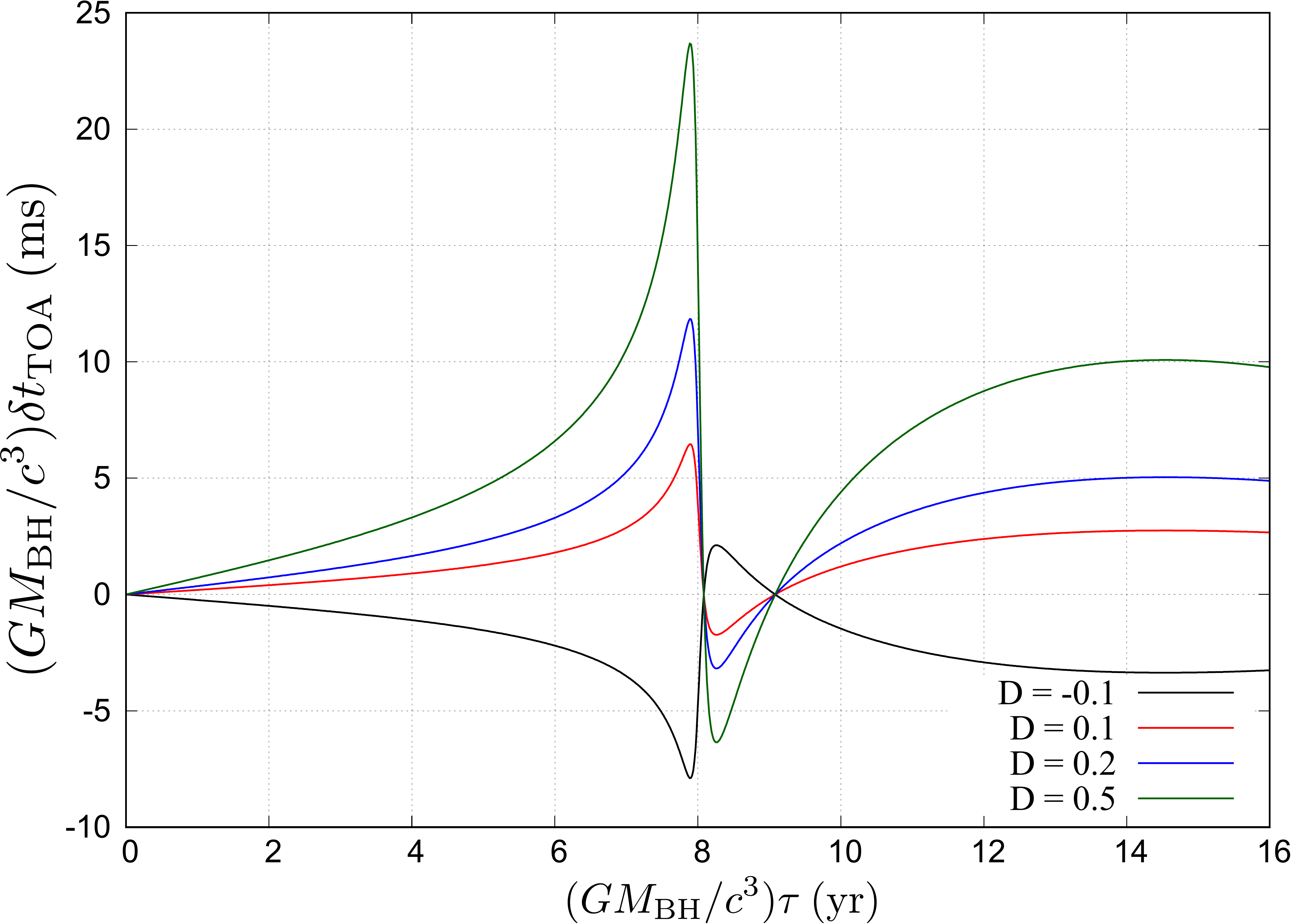}
\label{fig:devDKerr_Kerr_a07_right}
}
\end{center}
\caption{Differences in the apparent position (a) and the TOA (b) in the case
 $\tilde{a}_\ast=0.7$.
The solid lines with colors are the cases that $D=-0.1\,({\rm black}), 0.1\,({\rm red}), 
0.2\,({\rm blue})$, and $0.5\,({\rm green})$. (Color online)}
\label{fig:devDKerr_Kerr_a07}
\end{figure*}

We have examined three different spin parameters: $\tilde{a}_\ast=0.1$, $0.5$, and $0.7$.
The differences in the apparent position and the TOA are shown in figures \ref{fig:devDKerr_Kerr_a01}, 
\ref{fig:devDKerr_Kerr_a05} and \ref{fig:devDKerr_Kerr_a07}.
From figures \ref{fig:devDKerr_Kerr_a01_left}, \ref{fig:devDKerr_Kerr_a05_left}
and \ref{fig:devDKerr_Kerr_a07_left}, we find that $|\delta\alpha|$ and 
$|\delta\beta|$ are in the range from $10^{-3}$ to $10^{-2}\>\mu{\rm as}$.
From figures \ref{fig:devDKerr_Kerr_a01_right}, \ref{fig:devDKerr_Kerr_a05_right}
and \ref{fig:devDKerr_Kerr_a07_right}, 
we find that $|\delta t_{\rm TOA}|$ reaches the $10\>{\rm ms}$ order.
We also find that, although the disformal Kerr metric reduces to  
the Schwarzschild metric in the static limit with a finite value of $D$, since 
the possible range of $D$ is wider for smaller $\tilde{a}_\ast$, 
the effect of the disformal parameter $D$ can be more significant for a smaller value of $\tilde{a}_\ast$.


\subsubsection{Case II~(face-on: inclination angle $I=180^\circ$)}
\begin{figure*}[htbp]
\begin{center}
\subfigure[]
{
\includegraphics[width=70mm]{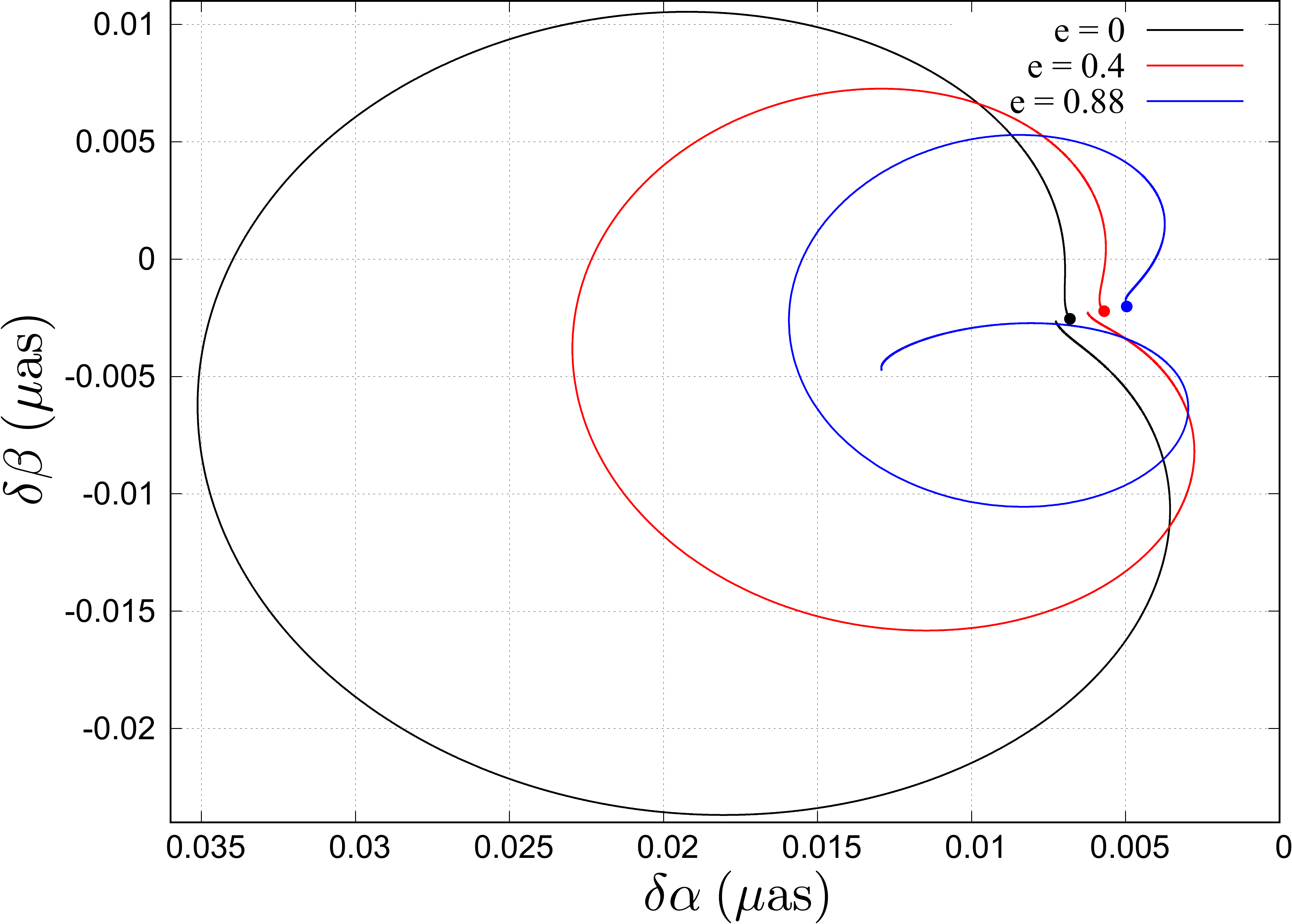}
\label{fig:devDKerr_Kerr_faceOn_left}
}
\subfigure[]
{
\includegraphics[width=70mm]{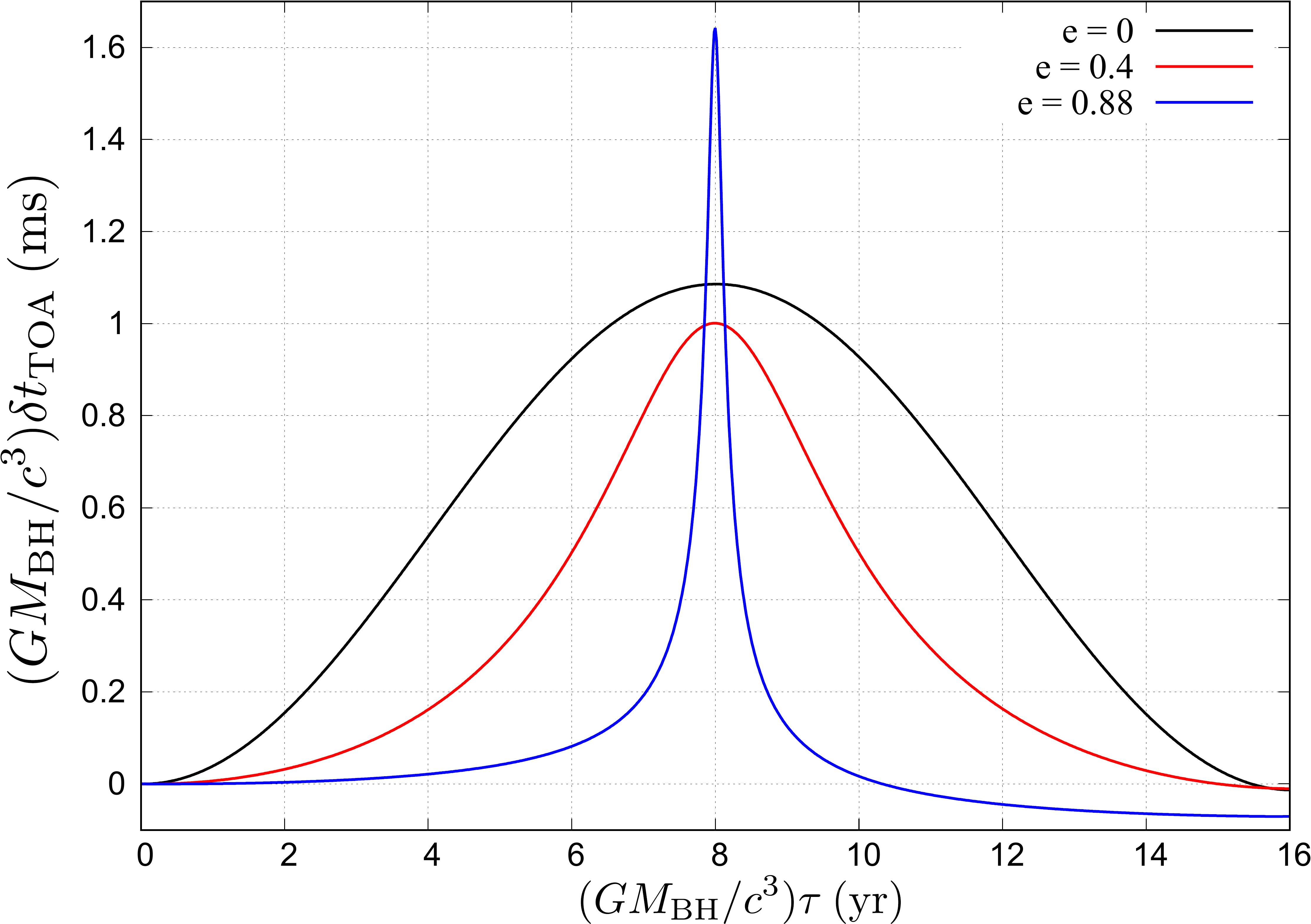}
\label{fig:devDKerr_Kerr_faceOn_right}
}
\end{center}
\caption{Differences in the apparent position (a) and the TOA (b) 
in the face-on case $I=180^{\circ}$.
The black, red, and blue solid lines show the cases that $e=0$, $0.4$, and $0.88$, respectively.
 (Color online)}
\label{fig:devDKerr_Kerr_faceOn}
\end{figure*}

The results are shown in figure \ref{fig:devDKerr_Kerr_faceOn}. 
We find that $|\delta\alpha|$ and $|\delta\beta|$ are the order of $10^{-2}\>\mu{\rm as}$.
The value of the apocenter shift is larger as the eccentricity increased.
The maximum values of $\delta t_{\rm TOA}$  are 
about $1\>{\rm ms}$ at around
the half of the period, as shown in figure \ref{fig:devDKerr_Kerr_faceOn_right}.

\subsubsection{Case III~(nearly edge-on: inclination angle $I=95^\circ$)}
\begin{figure*}[htbp]
\begin{center}
\subfigure[]
{
\includegraphics[width=70mm]{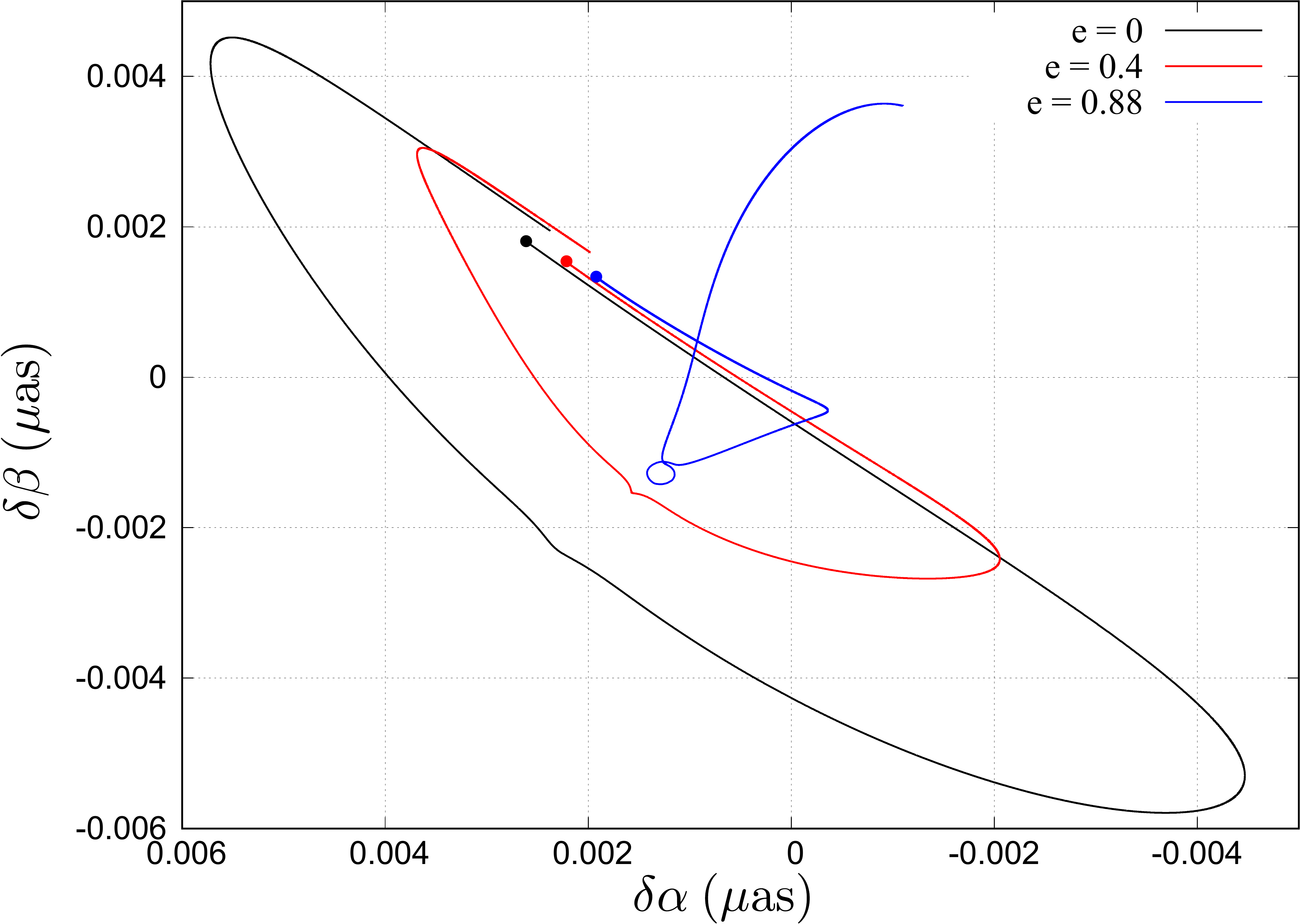}
\label{fig:devDKerr_Kerr_edgeOn_left}
}
\subfigure[]
{
\includegraphics[width=70mm]{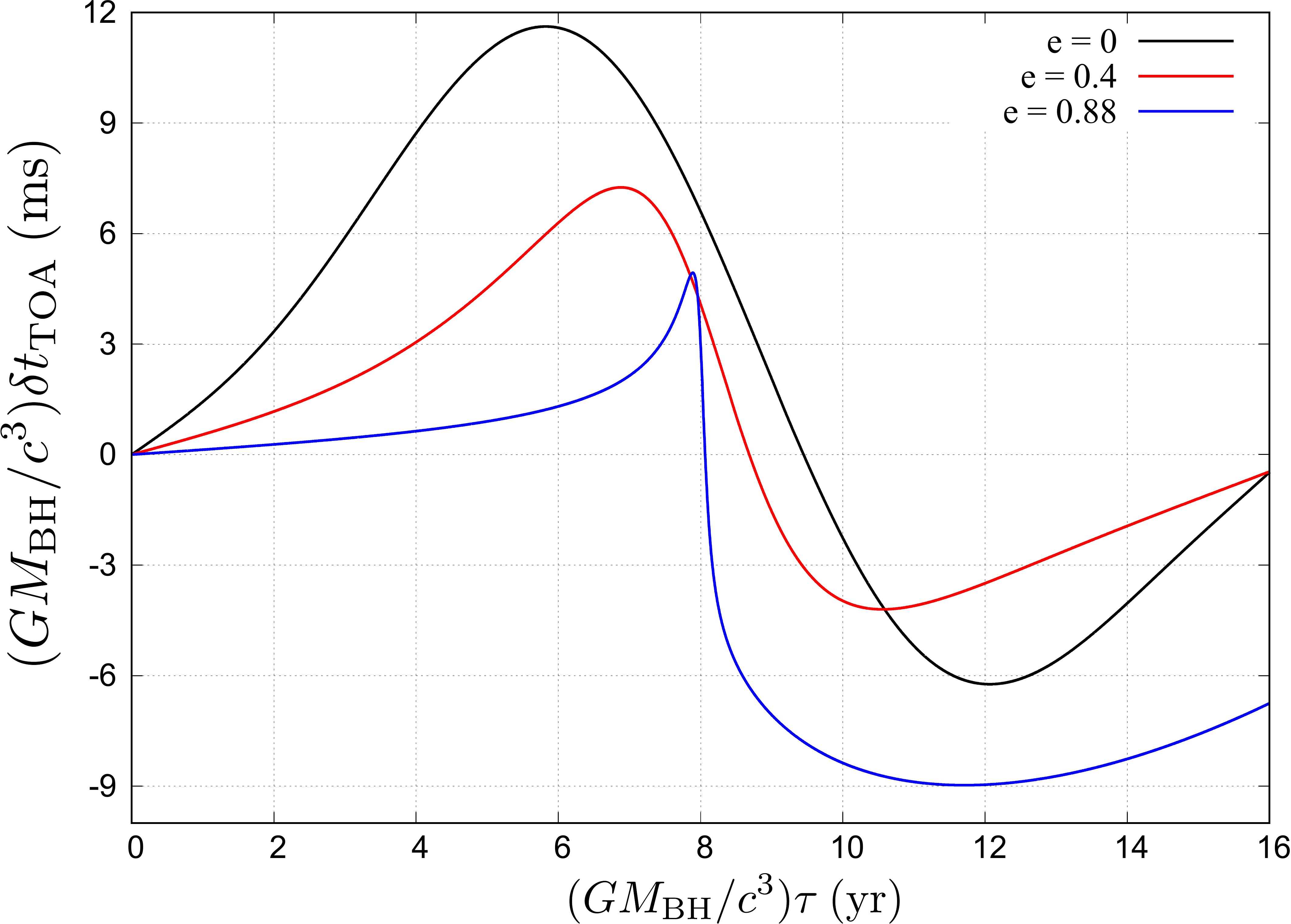}
\label{fig:devDKerr_Kerr_edgeOn_right}
}
\end{center}
\caption{Differences in the apparent position (a) and 
the TOA (b) in the nearly edge-on case $I=95^{\circ}$. 
The black, red, and blue solid lines show 
the cases that $e=0$, $0.4$, and $0.88$, respectively. (Color online)}
\label{fig:devDKerr_Kerr_edgeOn}
\end{figure*}

The results are shown in figure \ref{fig:devDKerr_Kerr_edgeOn}.
We find that $|\delta\alpha|$ and $|\delta\beta|$ are at the order of $10^{-3}\>\mu{\rm as}$, 
which is smaller than those of the face-on case.
The maximum value of $|\delta t_{\rm TOA}|$ can be $10\>{\rm ms}$ order,
which is greater than that of the face-on case. 
We also observe the significant apocenter shift for the case $e=0.88$.

\subsection{Detectability of the disformal parameter $D$ with the SKA}
We have calculated the apparent positions and the TOAs for each
S2-like, face-on, and nearly edge-on pulsar.
Here, we briefly discuss whether we can distinguish the disformal Kerr  
from the Kerr by observing such pulsars by the SKA.

First, let us focus on the apparent position.
Form figures \ref{fig:devDKerr_Kerr_a01_left},
 \ref{fig:devDKerr_Kerr_a05_left} and \ref{fig:devDKerr_Kerr_a07_left}, 
$|\delta\alpha|$ and $|\delta\beta|$ are in the range 
from $10^{-3}$ to $10^{-2}\>\mu{\rm as}$ per orbit.
In Case II and III, the differences in the apparent position are almost the same in Case I. 
The SKA could detect the apparent position of the pulsar with $10\>\mu{\rm as}$ accuracy.
Thus, we conclude that it is hard to detect the sign of 
the disformal Kerr black hole from the astrometric observation 
of pulsars on the orbit like S2 by the SKA.

Next, let us see the differences in the TOAs.
In Case I, from figures \ref{fig:devDKerr_Kerr_a01_right}, 
\ref{fig:devDKerr_Kerr_a05_right} and \ref{fig:devDKerr_Kerr_a07_right},
the maximum values of $|\delta t_{\rm TOA}|$ reach $10\>{\rm ms}$ order 
when $|D|\gtrsim 0.1$.
The accuracy of the TOA with the SKA is within $0.1$ to $10\>{\rm ms}$ for a normal pulsar, 
and therefore the differences in the TOA are sufficiently large for the detection. 
In Case II and III, we can find the typical magnitude of $\delta t_{\rm TOA}$ 
from figures \ref{fig:devDKerr_Kerr_faceOn_right} 
and \ref{fig:devDKerr_Kerr_edgeOn_right}.
Focusing on Case II (nearly edge-on), 
the maximum value of $|\delta t_{\rm TOA}|$
can be at the order of $10\>{\rm ms}$.
Since the value of $\delta t_{\rm TOA}$ is almost linearly dependent on the parameter $D$, 
we may conclude that the disformal parameter $|D|>0.1$ would be 
detectable with the observation of pulsars on the orbit like S2 by the SKA. 

\section{Summary and discussion}
Detecting possible deviations from the Kerr spacetime is an attractive  
subject to investigate a theory of gravity beyond general relativity. 
Recently, an exact solution called the disformal Kerr black hole solution has 
been constructed in a subclass of DHOST theories, and
the deviation from the Kerr solution is characterized by a constant parameter $D$.
In this paper, we have examined observational aspects of a hypothetical S2-like pulsar 
orbiting Sgr A* by assuming that Sgr A* is a disformal Kerr black hole. 
We numerically solved the equations of motion for
the S2-like pulsar and photons emitted from it with the Hamiltonian formalism. 
Then, we have shown how the disformal parameter $D$
affects the motion of the S2-like pulsar and the time of arrival of emitted photons.
Moreover, we discussed the detectability of the disformal parameter $D$ by 
a future radio telescope named the Square Kilometer Array (SKA).

First, we investigated a hypothetical S2-like pulsar orbiting 
a disformal Kerr black hole 
with the spin parameters $\tilde{a}_{\ast}=0.1, 0.5$, and $0.7$.
We found that the difference in 
the apparent position of the pulsar between the disformal Kerr and Kerr 
cases is about $10^{-2}\>\mu{\rm as}$ per orbit at most. 
It is smaller than the expected uncertainty of the SKA, that is $10\>\mu{\rm as}$.
For the emitted pulses, we estimated that the difference of the times of arrival between 
two cases can be at the order of $10\>{\rm ms}$ for $|D|\gtrsim 0.1$, 
which is sufficient for detection with the SKA.
The difference is mainly caused by the non-circularity of the disformal Kerr spacetime
appearing at $1.5$PN order.
Thus the SKA can probe the non-circularity of the spacetime around Sgr A* at $1.5$PN order.
Note that the allowed parameter range of the disformal parameter $D$ 
is restricted, depending on the spin parameter.
For the case of a rapidly rotating black hole, the allowed range of $D$ is 
small compared to the case of a slowly rotating black hole.
In the case of the spin parameter with $\tilde{a}_{\ast}\gtrsim 0.88$,  
the allowed disformal parameter is smaller than $0.1$.
We could not distinguish the disformal Kerr black hole
with $\tilde{a}_{\ast}\gtrsim 0.88$ from the Kerr spacetime by the SKA. 
Although no one knows the spin parameter for Sgr A* exactly,
it is estimated to be in the range $|\tilde{a}_{\ast}|\leq0.5$ 
\citep{2010MNRAS.403L..74K, 2020ApJ...901L..32F}.
Sgr A* can be a good candidate for the testing site of
the disformal Kerr black hole since the allowed range of $D$ is large 
for a slowly rotating black hole. 
Aside from a specific theory of modified gravity, since the non-circularity implies 
a deviation from the Kerr spacetime, one can perform an effective test 
of general relativity through the non-circularity of the spacetime around Sgr A* with the SKA.

Finally, we make a comment on the constraint obtained from gravitational wave observations.
It should be noted that, from the nearly simultaneous detection of the gravitational-wave and 
gamma-ray signals from a neutron-star merger event GW170817~\citep{2017PhRvL.119p1101A}, 
the difference between the speed of gravitational waves and that of light is 
tightly constrained~\citep{2018FrASS...5...44E}. 
In the DHOST theory, the speed of gravitational waves differs from the speed of 
light in general and the theory that we have used in this paper is not exceptional. 
Hence, the theory, at least a part of the theory, will not be viable on cosmological scales 
due to the above constraint.
Nevertheless, we hope that the black hole at the Galactic Center can provide a complementary 
test of modified gravity on different scales beside tests on cosmological scales.

%;Acknowledgment%%
%\bigskip
\begin{ack}
This work was supported by JSPS KAKENHI Grant Numbers JP19H00695, JP19H01900 (Y.~T.), 
20H05852 (A.N.), JP19H01891 (A.N. and D.Y.), 
17H01110, 20H00180, 21H01130, 21H04467 (K.T.), 17K14304 (D.Y.), 
JP19H01895, JP20H05850 and JP20H05853 (C.Y.), 
Bilateral Joint Research Projects of JSPS (K.T.), and the ISM Cooperative Research Program 
[2020-ISMCRP-2017] (K.T.).
\end{ack}

%;References%%


\begin{thebibliography}{}
\bibitem[Abbott et al.(2016)]{2016PhRvL.116f1102A} 
Abbott,~B.~P., et al. 2016, \prl, 116, 061102

\bibitem[Abbott et al.(2017)]{2017PhRvL.119p1101A} 
Abbott,~B.~P., et al. 2017, \prl, 119, 161101 

\bibitem[Achour et al.(2016)]{2016PhRvD..93l4005B} 
Achour,~J.~B., Langlois,~D., \& Noui,~K. 2016, \prd, 93, 124005

\bibitem[Achour et al.(2020)]{2020JCAP...11..001B} 
Achour,~J.~B., Liu,~H., Motohashi,~H., Mukohyama,~S., \& Noui,~K. 2020,
JCAP, 11, 001
%Journal of Cosmology and Astroparticle Physics {\bf 11}, 001 (2020). 

\bibitem[Ang{\'e}lil \& Saha(2010)]{2010ApJ...711..157A} 
Ang{\'e}lil,~R., \& Saha,~P. 2010, \apj, 711, 157

\bibitem[Anson et al.(2021a)]{2021PhRvD.103l4035A}
Anson,~T., Babichev,~E., \& Charmousis,~C. 2021a, \prd, 103, 124035

\bibitem[Anson et al.(2021b)]{2021JHEP...01..018A} 
Anson,~T., Babichev,~E., Charmousis,~C., \& Hassaine,~M. 2021b, 
JHEP, 2021, 18
%Journal of High Energy Physics

\bibitem[Bates et al.(2011)]{2011MNRAS.411.1575B} 
Bates,~S.~D., et al. 2011, \mnras, 411, 1575

\bibitem[Charmousis et al.(2019)]{2019PhRvD.100h4020C} 
Charmousis,~C., Crisostomi,~M., Gregory,~R., \& Stergioulas,~N. 2019,
\prd, 100, 084020

\bibitem[Chen et al.(2021)]{2021arXiv210311788C}
Chen,~S., Wang,~Z., \& Jing.~J. 2021, %arXiv:2103.11788 (2021).
JCAP 06, 043

\bibitem[Deneva et al.(2009)]{2009ApJ...702L.177D} 
Deneva,~J.~S., Cordes,~J.~M., \& Lazio,~T.~J.~W. 2009, \apjl, 702, L177

\bibitem[Do et al.(2019)]{2019Sci...365..664D} 
Do,~T., et al. 2019, Science, 365, 664. 

\bibitem[Eatough et al.(2013)]{2013Natur.501..391E}
Eatough,~R.~P., et al. 2013, \nat, 501, 391

\bibitem[EHT Collab.(2019)]{2019ApJ...875L...1E} 
Event Horizon Telescope Collaboration 2019, \apjl, 875, L1

\bibitem[Ezquiaga \& Zumalac{\'a}rregui(2018)]{2018FrASS...5...44E} 
Ezquiaga,~J.~M., \& Zumalac{\'a}rregui,~M. 2018, 
%Frontiers in Astronomy and Space Sciences, 5, 44 
Front. Astron. Space Sci., 5, 44

\bibitem[Fragione \& Loeb(2020)]{2020ApJ...901L..32F} 
Fragione,~G., \& Loeb,~A. 2020, \apjl,  901, L32

\bibitem[Genzel et al.(2010)]{2010RvMP...82.3121G} 
Genzel,~R., Eisenhauer,~F., \& Gillessen,~S. 2010, 
%Reviews of Modern Physics
Rev. Mod. Phys., 82, 3121

\bibitem[GRAVITY Collab.(2018)]{2018AA...615L..15G}
GRAVITY Collaboration 2018, \aap, 615, L15 

\bibitem[GRAVITY Collab.(2020)]{2020AA...636L...5G} 
GRAVITY Collaboration 2020,  \aap,  636, L5

\bibitem[Kato et al.(2010)]{2010MNRAS.403L..74K} 
Kato,~Y., Miyoshi,~M., Takahashi,~R., Negoro,~H., \& Matsumoto,~R. 2010,
\mnras, 403, L74

\bibitem[Kennea et al.(2013)]{2013ApJ...770L..24K} 
Kennea,~J.~A., et al. 2013, \apjl,  770, L24

\bibitem[Kobayashi(2019)]{2019RPPh...82h6901K} 
Kobayashi,~T. 2019, Rep. Prog. Phys., 82, 086901 

\bibitem[Kramer et al.(2006)]{2006Sci...314...97K} 
Kramer,~M., et al. 2006, Science, 314, 97

\bibitem[Langlois \& Noui(2016)]{2016JCAP...02..034L} 
Langlois,~D., \& Noui,~K. 2016, 
%Journal of Cosmology and Astroparticle Physics 
JCAP, 2016, 034

\bibitem[Langlois(2019)]{2019IJMPD..2842006L}
Langlois,~D. 2019, Int. J. Mod. Phys. D, 28, 1942006-3287 

\bibitem[Liu et al.(2021)]{2021ApJ...914...30L} 
Liu,~K., et al. 2021,  \apj, 914, 30

\bibitem[Liu et al.(2012)]{2012ApJ...747....1L} 
Liu,~K., Wex,~N., Kramer,~M., Cordes,~J.~M., \& Lazio,~T.~J.~W. 2012,
\apj, 747, 1

\bibitem[Macquart et al.(2010)]{2010ApJ...715..939M} 
Macquart,~J.~P., Kanekar,~N., Frail,~D.~A., \& Ransom,~S.~M. 2010, 
\apj, 715, 939

\bibitem[Mori et al.(2013)]{2013ApJ...770L..23M} 
Mori,~K., et al. 2013, \apjl, 770, L23

\bibitem[Nakashi \& Kimura(2020)]{2020PhRvD.102h4021N} 
Nakashi,~K., \& Kimura,~M. 2020, \prd, 102, 084021

\bibitem[Pfahl \& Loeb(2004)]{2004ApJ...615..253P} 
Pfahl,~E., \& Loeb,~A. 2004, \apj, 615, 253

\bibitem[Psaltis et al.(2016)]{2016ApJ...818..121P} 
Psaltis,~D., Wex,~N., \& Kramer,~M. 2016, \apj, 818, 121 

\bibitem[Robinson(1975)]{1975PhRvL..34..905R} 
Robinson,~D.~C. 1975, \prl, 34, 905 

\bibitem[Smits et al.(2009)]{2009AA...493.1161S} 
Smits,~R., Kramer,~M., Stappers,~B., Lorimer,~D.~R., Cordes,~J., \& Faulkner,~A. 2009,
\aap, 493, 1161

\bibitem[Torne et al.(2021)]{2021AA...650A..95T} 
Torne,~P., et al. 2021, \aap, 650, A95

\bibitem[Wex \& Kopeikin(1999)]{1999ApJ...514..388W}
Wex,~N., \&  Kopeikin,~S.~M. 1999, \apj, 514, 388

\bibitem[Wharton et al.(2012)]{2012ApJ...753..108W} 
Wharton,~R.~S., Chatterjee,~S., Cordes,~J.~M., Deneva,~J.~S., \& 
Lazio,~T.~J.~W. 2012, \apj,  753, 108

\bibitem[Xie et al.(2021)]{2021PhRvL.126x1104X} 
Xie,~Y., Zhang,~J., Silva,~H.~O., Rham,~C., Witek,~H., \& Yunes,~N. 2021,
\prl, 126, 241104

\bibitem[Zhang et al.(2015)]{2015ApJ...809..127Z} 
Zhang,~F., Lu,~Y., \& Yu,~Q. 2015, \apj, 809, 127

\bibitem[Zhang \& Saha(2017)]{2017ApJ...849...33Z} 
Zhang,~F., \& Saha,~P. 2017, \apj, 849, 33

\end{thebibliography}
\end{document}